\documentclass[12pt]{article}
\usepackage{bbm}
\usepackage{comment}

\usepackage{subfig}
\usepackage{graphicx}
\usepackage{caption}
\usepackage{changepage}
\usepackage{bm, multirow, multicol}
\usepackage{caption}
\usepackage{float}
\usepackage{amssymb}
\usepackage{amsmath}
\usepackage{algorithm}
\usepackage{hyperref}
\usepackage[noend]{algpseudocode}
\usepackage{placeins} 
\usepackage{xcolor}
\makeatletter
\def\BState{\State\hskip-\ALG@thistlm}

\newcommand{\RNum}[1]{\uppercase\expandafter{\romannumeral #1\relax}}
\makeatother
\usepackage{bm}

\usepackage{authblk}
\usepackage[margin=.75in]{geometry}
\usepackage{color}

\definecolor{mygreen}{RGB}{120, 118, 56}

\graphicspath{{./figures/}{./figures/sampling_validation/}}

\title{\vspace*{-1.3cm}On the role of fractional Brownian motion in models of chemotaxis and stochastic gradient ascent}
\author[1]{Gustavo Cornejo-Olea}
\author[1]{Lucas Buvinic}
\author[2]{J\'er\^ome Darbon}
\author[3]{Radek Erban}
\author[4]{\\Andrea Ravasio}
\author[1]{Anastasios Matzavinos}
\affil[1]{\normalsize Institute for Mathematical and Computational Engineering, Pontifical Catholic University of Chile, Santiago, Chile}
\affil[2]{\normalsize Division of Applied Mathematics, Brown University, Providence, RI 02912, USA}
\affil[3]{\normalsize Mathematical Institute, University of Oxford, Andrew Wiles Building, Radcliﬀe Observatory
Quarter, Woodstock Road, Oxford OX2 6GG, United Kingdom}
\affil[4]{\normalsize Institute for Biological and Medical Engineering, Pontifical Catholic University of Chile, Santiago, Chile}
\date{}

\begin{document}
\maketitle
\vspace*{-0.7cm}\noindent\textbf{Abstract}:  Cell migration often exhibits long-range temporal correlations and anomalous diffusion, even in the absence of external guidance cues such as chemical gradients or topographical constraints.  This raises a fundamental question: do such correlations enhance a cell's ability to navigate complex environments? In this work, we explore how temporally correlated noise (modeled using fractional Brownian motion) influences chemotactic search dynamics. Through computational experiments, we show that superdiffusive motion, when combined with gradient-driven migration, enables robust exploration of the chemoattractant landscape. Cells reliably reach the global maximum of the concentration field, even in the presence of spatial noise, secondary cues, or irregular signal geometry. We quantify this behavior by analyzing the distribution of first hitting times under varying degrees of temporal correlation. Notably, our results are consistent across diverse conditions, including flat and curved substrates, and scenarios involving both primary and self-generated chemotactic signals. Beyond biological implications, these findings also offer insight into the design of optimization and sampling algorithms that benefit from structured stochasticity.\bigskip
\smallskip

\section{Introduction}\label{intro}

The directed movement of cells and microorganisms in response to a diffusible chemical signal is known as chemotaxis \cite{RE:2004,RE:2005,RE:2007,chem:2016}. The first mathematical model of chemotaxis was introduced by Evelyn Keller and Lee Segel, who proposed a parabolic system of partial differential equations to study the aggregation dynamics of the social amoeba \textit{Dictyostelium discoideum} \cite{KS70,KS71}. Since then, the Keller--Segel model has been studied extensively; a comprehensive overview of related mathematical developments is provided in \cite{Perthame:2007}. Several extensions of the Keller--Segel system at the population level have been proposed, including nonlocal models \cite{Hillen,matzavinos2}, partial differential equation models posed in random heterogeneous media \cite{matzavinos1}, and stochastic partial differential equations \cite{Chen:2025,Huang:2021}.

At the single-cell level, stochastic differential equation models of chemotaxis are also widely used. These include particle-based approximations of the classical Keller--Segel system \cite{Schmeiser:2009,Schmeiser:2011,AStevens:2000,Zhang:2025}, as well as computational models of various biological and pathophysiological processes involving chemotactic motility \cite{RE:2025,RE:2020}. Examples include granulocyte migration \cite{granulocyte}, cancer invasion and metastasis \cite{Katsaounis}, tumor-induced angiogenesis \cite{Capasso:2019,Capasso,CLStokes}, and cell aggregation and self-organization phenomena \cite{Chalub:2006,ABC:2025}. These models often combine Brownian motion with gradient-driven motility, drawing parallels to stochastic gradient ascent algorithms used in global optimization \cite{Matus:2017,Blei:2017} and sampling methods \cite{Durmus}. Viewed through this lens, chemotaxis can be interpreted as a natural optimization process that steers cells toward regions of higher chemoattractant concentration.

However, the assumption that random motility follows Brownian dynamics -- and thus exhibits normal diffusion -- does not always align with experimental observations. In the absence of external guidance cues such as chemotactic gradients or topographical constraints, the movement of biological cells often deviates from Brownian motion. A growing body of evidence shows that single-cell trajectories frequently exhibit \emph{anomalous diffusion}, characterized by a mean squared displacement (MSD) that scales nonlinearly with time. Such behavior has been reported across diverse cell types, including fibroblasts \cite{Selmeczi2005}, epithelial cells \cite{Dieterich2008}, amoeboid cells such as \textit{Dictyostelium} \cite{Makarava:2014,Takagi2008}, and immune cells \cite{Dieterich2022, Harris2012}, both {\it in vitro} and {\it in vivo}. Depending on the underlying mechanisms, the random motility may be \emph{subdiffusive} or \emph{superdiffusive}, often reflecting cytoskeletal dynamics, intermittent propulsion, or memory in the direction of motion.

A widely used stochastic model for capturing such correlated motion is \emph{fractional Brownian motion} (fBM) \cite{Nourdin:2012,Mishura:2017}, a self-similar Gaussian process with stationary increments and long-range temporal correlations -- see Section \ref{sct:fbm} for the mathematical definition. Several studies have successfully fitted fBM models to cell migration data, using metrics such as MSD scaling, velocity autocorrelations, and trajectory segmentation to infer time-dependent persistence \cite{Korabel2022,Makarava:2014}. 
In particular, the anomalous motility of epithelial cells \cite{Dieterich2008}, neutrophils \cite{Dieterich2022}, and Drosophila hemocytes \cite{Korabel2022} has been quantitatively described using fBM or related fractional models. These observations suggest that memory effects in spontaneous cell migration -- present even in the absence of external cues -- may play a critical role in tissue-scale pattern formation.

Motivated by these observations, this work explores how fractional Brownian motion shapes the dynamics of chemotactic cell migration. Specifically, we ask whether the long-range temporal correlations reported in experimental studies can improve a cell's ability to navigate toward a chemoattractant source.

Through computational experiments, we show that superdiffusive random motion, when combined with gradient-driven migration, promotes robust exploration of the chemoattractant landscape. In this regime, small fluctuations in concentration do not prevent cells from locating regions of higher signal intensity. To quantify this effect, we analyze the distribution of first hitting times for the global maximum of the chemoattractant concentration over fixed observation windows. Notably, this behavior persists across diverse settings, including flat and curved substrates, and in the presence or absence of secondary, self-generated chemotactic cues.

The structure of the paper is as follows. Section 2 introduces the mathematical definition and key properties of fractional Brownian motion, and sets up the stochastic differential equation model of chemotaxis that underlies our analysis. This framework includes dynamics in both Euclidean spaces and Riemannian manifolds, representing flat and curved substrates, respectively. We also discuss connections to optimization and sampling theory. Section 3 presents our computational results: Section 3.1 examines the robustness conferred by fractional Brownian motion against small fluctuations in the chemotactic field in Euclidean settings; Section 3.2 extends this analysis to curved domains modeled as smooth manifolds; and Section 3.3 investigates the influence of a secondary, self-generated chemotactic cue. Finally, Section 4 concludes with a discussion of our findings and directions for future research.

\section{Mathematical framework}\label{mm}

\subsection{Fractional Brownian motion and chemotaxis}\label{sct:fbm}

Fractional Brownian motion (fBM) was first introduced by Kolmogorov in the 1940s in the context of his mathematical studies of turbulence in fluid dynamics \cite{Kolmogorov}. The process was later revisited by Mandelbrot and Van Ness in the late 1960s with a view toward a broader range of applications, including economics, solid mechanics, and hydrology \cite{VanNess}. Their work also introduced the modern terminology and formal definition of fractional Brownian motion that remain in use today \cite{Mishura:2017,Mishura:2008,Nourdin:2012}.

Formally, fractional Brownian motion with Hurst index $H \in (0,1)$ is a centered Gaussian process $\left(W^H_t\right)_{t \geq 0}$ with continuous paths and stationary increments, characterized by the relation
\begin{equation}\label{incr}
\mathbb{E} \left[\bigl(W^H_t-W^H_s\bigr)^2\right] =  |t - s|^{2H}
\end{equation}
for all $t,s \geq 0$. It follows immediately that when $H = 1/2$, fractional Brownian motion coincides with standard Brownian motion. Moreover, according to \eqref{incr}, the process exhibits superdiffusive behavior when $H > 1/2$ and subdiffusive behavior when $H < 1/2$.

A straightforward algebraic manipulation of \eqref{incr} yields that for all $H \in (0,1)$,
\begin{equation}
\mbox{Cov} \bigl( W^H_t, W^H_s\bigr) = \mathbb{E}\bigl[ W^H_tW^H_s \bigr] = \frac{1}{2}\left( t^{2H} + s^{2H} - \left| t - s \right|^{2H} \right)
\label{eqn: covariance-fbm}
\end{equation}
Assuming $t > s > 0$ and that $W^H_0 = 0$, we then obtain
\begin{equation}\label{incr2}
\begin{array}{l}
\displaystyle \mathbb{E} \left[\bigl(W^H_s-W^H_0\bigr)\bigl(W^H_t-W^H_s\bigr)\right] = \mathbb{E}\bigl[ W^H_t W^H_s \bigr] - \mathbb{E}\bigl[ (W^H_s)^2\bigr]\medskip\\
\displaystyle \phantom{\mathbb{E} \left[\bigl(W^H_s-W^H_0\bigr)\bigl(W^H_t-W^H_s\bigr)\right]} = \frac{1}{2} \left( t^{2H} - s^{2H} - \left| t - s \right|^{2H} \right)
\end{array}
\end{equation}
Note that the right-hand side of \eqref{incr2} is positive when $H > 1/2$ and negative when $H < 1/2$. Therefore, the increments of fractional Brownian motion are positively correlated when $H > 1/2$ and negatively correlated when $H < 1/2$.

As discussed in the introduction, in this paper we are concerned with cell motility models in which the position $\mathbf{X}_t$ of a motile cell is driven by a combination of the gradient of a chemoattractant and a fractional Brownian motion with Hurst index $H$. Accordingly, $\mathbf{X}_t$ satisfies the following initial value problem:
\begin{equation}\label{SDE1}
\left\{
\begin{aligned}
d_{}\mathbf{X}_{t} &= \chi(\mathbf{X}_t)\nabla f(\mathbf{X}_t)\,dt \,+ \mathbf{\Sigma}(\mathbf{X}_t)\,d_{}\mathbf{W}_t^H \\
\mathbf{X}_0 &= \mathbf{x}_0
\end{aligned}
\right.
\end{equation}
Here, for all $t \geq 0$, the function $\chi(\,\cdot\,)$ is non-negative and represents chemotactic sensitivity, while $\mathbf{\Sigma}(\,\cdot\,)$ denotes an $n \times n$ noise amplitude matrix. The process $\mathbf{W}_t^H$ is a vector-valued fractional Brownian motion, whose components are independent scalar fractional Brownian motions with common Hurst index $H$. When $H = 1/2$, the stochastic differential $d_{}\mathbf{W}_t^H$ corresponds to white noise. In this work, we focus on the case $H > 1/2$, which corresponds to noise with positive temporal autocorrelation~\cite{Mishura:2017}.

Equation \eqref{SDE1} is typically interpreted in its weak form as the integral equation
\begin{equation}\label{SDE2}
\mathbf{X}_t = \mathbf{x}_0 + \int_0^t  \chi(\mathbf{X}_s)\nabla f(\mathbf{X}_s)_{}  \,ds + \int_0^t \mathbf{\Sigma}(\mathbf{X}_s)\,d_{}\mathbf{W}_s^H
\end{equation}
The definition of the stochastic integral in Eq.~\eqref{SDE2} depends on the value of the Hurst index $H$. When $H < 1/2$, the integral is typically defined using tools from Malliavin calculus \cite{Decreusefond}. When $H = 1/2$, Eq.~\eqref{SDE2} is interpreted in the sense of the It\^o integral \cite{Oksendal}. In the case of interest for this paper, $H > 1/2$, one can take advantage of the pathwise regularity of $\mathbf{W}^H_t$ to define the stochastic integral as a Young integral; see \cite{Nourdin:2012} for further details. A well-developed theoretical framework ensures the existence and uniqueness of solutions to Eq.~\eqref{SDE2}; see \cite{Nourdin:2012,Mishura:2008}.

For simplicity, most of the computational experiments presented in Section~\ref{results} assume that $\chi(\,\cdot\,)$ is constant, i.e., $\chi(\,\cdot\,) \equiv \chi > 0$, and that the noise amplitude matrix is a scalar multiple of the identity matrix, i.e., $\mathbf{\Sigma}(\,\cdot\,) = \sigma\,\mathbf{I}$ for some $\sigma > 0$. In the experiments of Sections~3.1 and~3.2, these coefficients remain of this form only while the local chemoattractant concentration stays below a prescribed threshold. Once the threshold is exceeded, the motility program is terminated, so that both the chemotactic response and the random exploratory component are switched off. This is intended to model the cell ceasing exploratory motion and becoming localized near the chemoattractant source after successful detection. All numerical simulations have been performed with an Euler-type scheme; see, e.g., \cite{Mishura_p2008}.

\subsection{Chemotaxis on a manifold}\label{manifold}

In many pathophysiological settings, chemotaxis occurs on curved substrates, suggesting that curvature may influence cell motility dynamics \cite{cornea}.
 To account for this, we model the substrate as a smooth Riemannian manifold $(\mathcal{M}, g)$ with metric tensor $g = (g_{ij})$ \cite{Boumal,Chow2}, and we provide the corresponding local coordinate expressions for Eq.~\eqref{SDE1}, assuming all processes evolve on the manifold. For the Brownian motion case, these expressions can be found in \cite{Emery,Hsu,RogWil}, while for fractional Brownian motion with Hurst index $H > 1/2$, the derivation is provided in Appendix 1.

Specifically, consider a chemotactic process $\mathbf{Y}_{t}$ on a Riemannian manifold $(\mathcal{M}, g)$, driven by the intrinsic gradient of a smooth function $f : \mathcal{M} \to \mathbb{R}$ and by a fractional Brownian motion $\mathbf{B}^{H}_t$ defined on the manifold.  The evolution of $\mathbf{Y}_{t}$ is then governed by

\begin{equation}\label{SDE_m}
\left\{
\begin{aligned}
d_{} \mathbf{_{}Y}_{t} &= \chi\,\operatorname{grad}_{\mathcal{M}}f(_{}\mathbf{Y}_{t})\,dt \,+ \sigma\,d_{}\mathbf{B}_t^H \\
 \mathbf{_{}Y}_0  &= \mathbf{y}_0 \in \mathcal{M}
\end{aligned}
\right.
\end{equation}

Similarly to Section \ref{sct:fbm}, the weak formulation of Eq. \eqref{SDE_m} depends on the value of the Hurst index $H$, as we now explain. For the sake of simplicity, we assume that $\mathcal{M}\subset\mathbb{R}^N$ is a smooth embedded $n$-dimensional manifold, with a global chart
\begin{equation}\label{glch}
\varphi: U\subset\mathbb{R}^n \;\longrightarrow\; \mathcal{M}\subset\mathbb{R}^N
\end{equation}
Of course, in this paper we will mainly focus on the biologically relevant case of $n=2$ and $N=3$. As usual, $U$ denotes an open subset of $\mathbb{R}^n$.

Let us recall the definition of fractional Brownian motion on a Riemannian manifold. 
For $H=1/2$, Brownian motion on $\mathcal{M}$ is classically defined as the 
{\it Stratonovich} solution to the stochastic differential equation
$$
\partial_{}\mathbf{B}_t = \sum_{k=1}^n E_k\bigl(\mathbf{B}_t\bigr)\,\partial_{}W_t^{k}, \qquad \mathbf{B}_0 \in \mathcal{M},
$$
where $\{E_k\}_{k=1}^n$ is a local orthonormal frame of tangent vector fields on $\mathcal{M}$ 
and $\{W^k_t\}_{k=1}^n$ are independent (one-dimensional, Euclidean) Brownian motions \cite{RogWil}. 
This formulation ensures that $\mathbf{B}_t\in\mathcal{M}$ for all $t\geq 0 $ and that the generator of the process is given by
$\tfrac12\Delta_g$, where $\Delta_g$ is the Laplace-Beltrami operator on $\mathcal{M}$ \cite{Hsu,RogWil}.  

When $H>1/2$, the same equation can be interpreted in the {\it Young sense}, with $\{W^{H,k}_t\}_{k=1}^n$ 
independent fractional Brownian motions of Hurst index $H$.  In this case, the manifold-valued process 
$\mathbf{B}_t^H$ is defined as the solution to
\begin{equation}\label{frame_def}
d_{}\mathbf{B}_t^H = \sum_{k=1}^n E_k\bigl(\mathbf{B}_t^H\bigr)\, d_{}W_t^{H,k}, 
\qquad \mathbf{B}_0^H \in \mathcal{M},
\end{equation}
where all stochastic differentials are understood in the Young sense \cite{Nourdin:2012}.  
Equation~\eqref{frame_def} generalizes the Stratonovich formulation of Brownian motion to the fractional setting and yields a process whose increments are isotropic in the tangent space of $\mathcal{M}$.  
In particular, when $\mathcal{M}$ is a Lie group equipped with a left-invariant metric and $\{E_k\}$ are the corresponding left-invariant orthonormal vector fields, \eqref{frame_def} coincides with the definition of fractional Brownian motion introduced in~\cite{BaudoinCoutin}.

In the presence of the global chart $\varphi$ given in \eqref{glch}, one can express Eq.~\eqref{SDE_m} in local coordinates $\{x^i\}_{i \leq n}$ as follows. Using the notation $\partial_k = \partial/\partial x^k$, the $k$-th component of the gradient of $f$ can be expressed  as
$$
(\operatorname{grad}_{\mathcal{M}}f)^k = \sum_{j=1}^n g^{kj}\,\partial_j f, \quad k = 1, \dots, n, 
$$
where, as usual, $g^{kj}$ denotes the $(k,j)$ entry of the {\it inverse} metric tensor \cite{Boumal,Chow2}. Note that throughout this paper, we use the commonly adopted convention for raised and lower indices, representing components of contravariant and covariant tensors, respectively \cite{Chow2}.  

As explained in detail in Appendix 1, one can also express $\mathbf{B}^H_t$ in local coordinates, which in turn yields a stochastic differential equation for the evolution of the process 
$$\mathbf{X}_t=\varphi^{-1}(\mathbf{Y}_t)\in U$$ 
Indeed, let $W^{H,j}_t$ denote the (one-dimensional) $j$-th component of a Euclidean $n$-dimensional fractional Brownian motion $\mathbf{W}^H_t$ with Hurst index $H\geq 1/2$, and let $X^k_t$ denote the $k$-th component of the local-coordinate process $\mathbf{X}_t$.  Then, the chemotactic process $\mathbf{Y}_t$ on $\mathcal{M}$ given by \eqref{SDE_m} satisfies the following system of stochastic differential equations in local (Euclidean) coordinates:
\begin{align}\label{SDE:local} 
d X_t^k = \left(\chi\, \sum_{j=1}^n g^{kj}\,\partial_j f - \frac{\sigma^{2\,} \mathbbm{1}_{\{H=1/2\}}}{2} \sum_{i,j=1}^n g^{ji}\, \Gamma^k_{ij}\right) dt + \sigma \sum_{j=1}^n
\bigl(g^{-1/2\,}\bigr)^k_j\,dW_t^{H,j}, \quad k = 1, \dots, n,
\end{align}
which is to be interpreted in the It\^o sense when $H=1/2$ and in the Young sense when $H>1/2$, and $\Gamma^k_{ij}$ are the Christoffel symbols of the Levi-Civita connection on $\mathcal{M}$ \cite{Chow2,Hsu}.

Note the appearance of an additional advection term when $H=1/2$. This is due to the transformation of the Stratonovich formulation to the It\^o formulation in local coordinates (see, e.g., Example 3.3.5 in \cite{Hsu}). In contrast, the Young integral behaves like the Stratonovich integral in terms of change of coordinates \cite{Mishura:2017,Nourdin:2012}, and hence the additional advection term is absent when $H>1/2$; see Appendix 1 for more details. Thus, the motility parameters in \eqref{SDE:local} are influenced by the inverse metric tensor and the Christoffel symbols of $\mathcal{M}$. In Section~\ref{res:man}, we will examine how the geometry of $\mathcal{M}$ affects the motility dynamics of a cell performing chemotaxis on a curved substrate.

\section{Results}\label{results}
\subsection{Robustness against signal fluctuations in Euclidean domains}\label{res:Euclid}
\begin{figure}[!t]
\begin{center}
\subfloat{
    \includegraphics[width=0.48\textwidth]{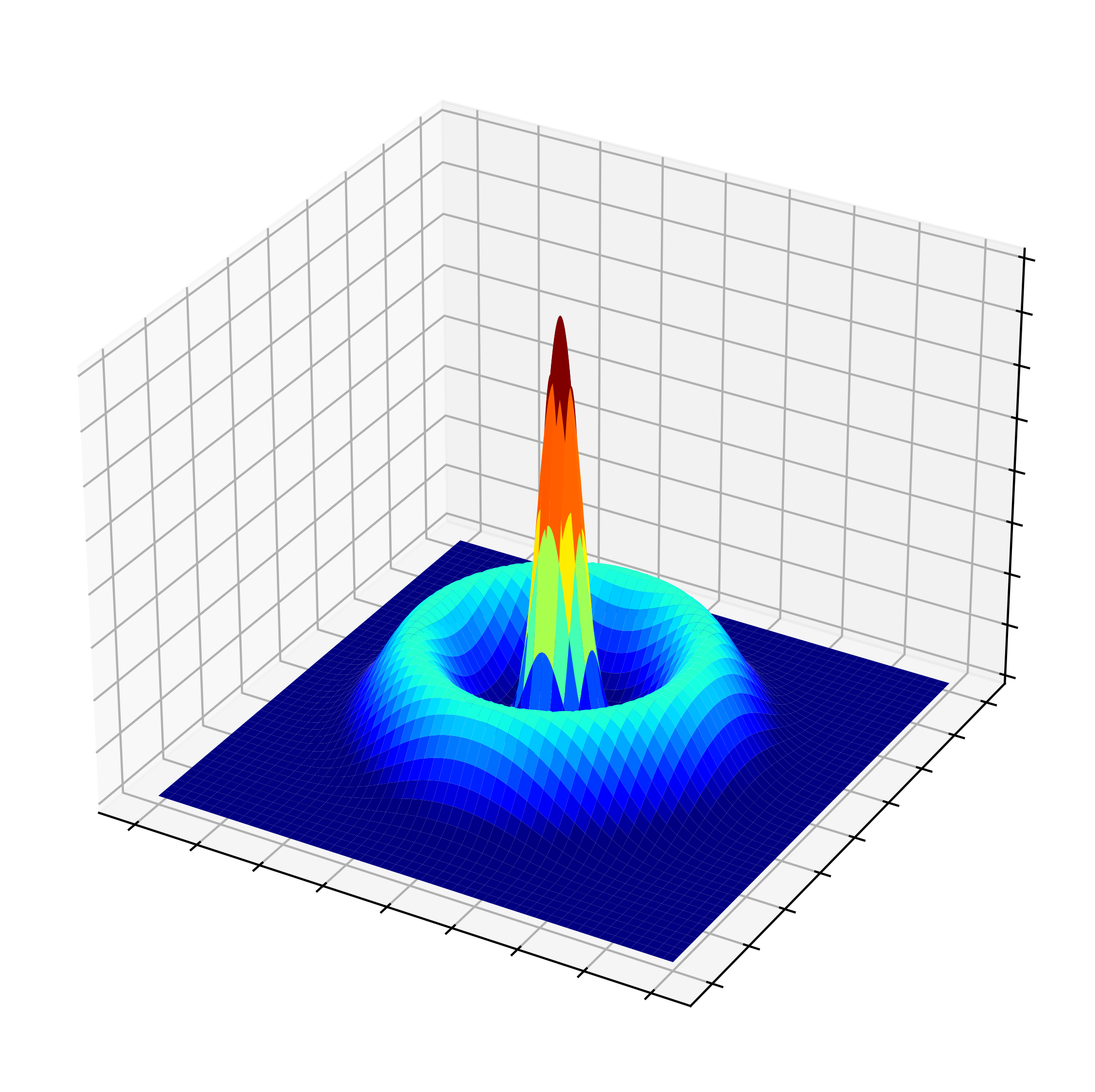}
}
\subfloat{
    \includegraphics[width=0.49\textwidth]{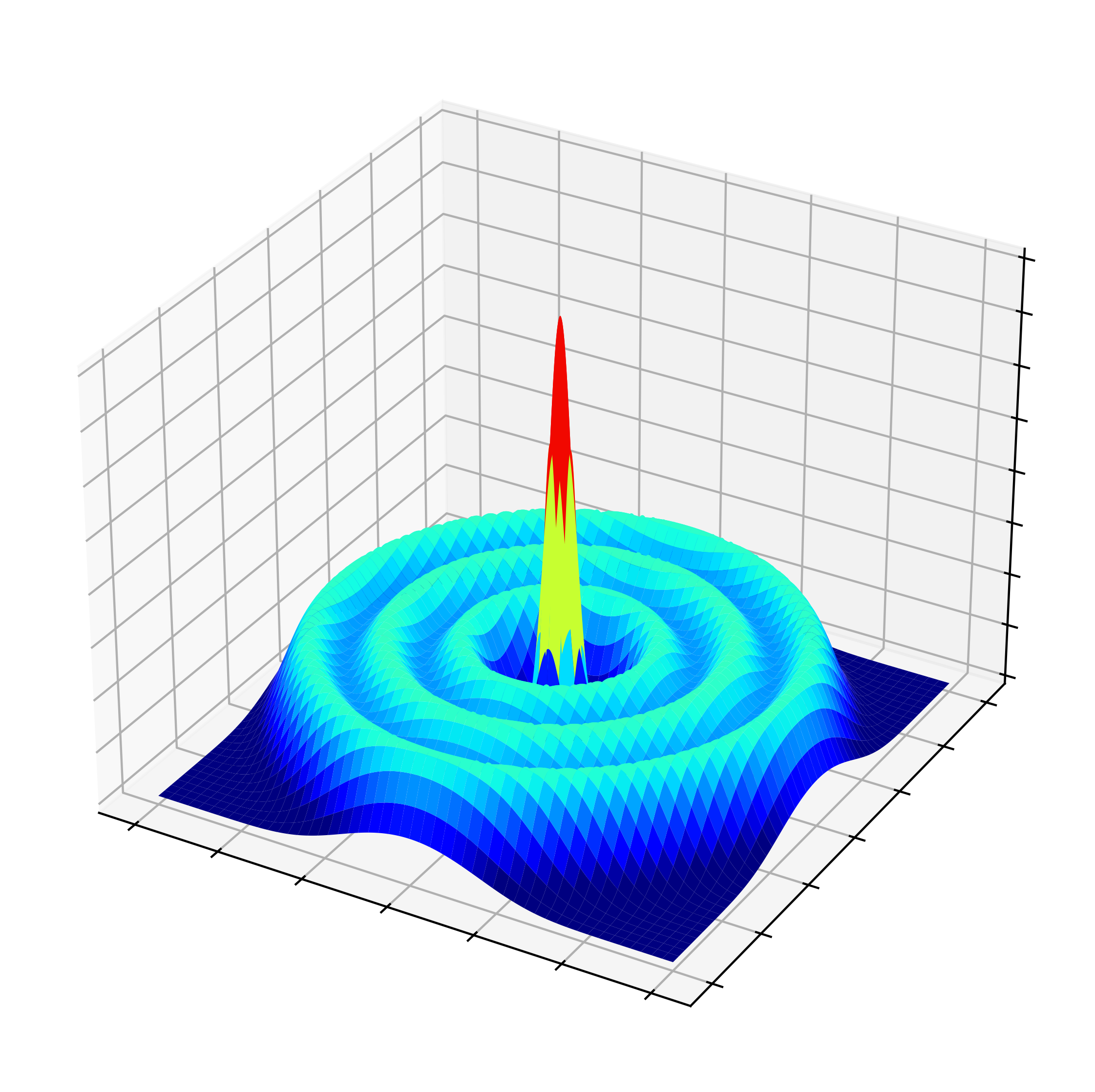}
}
\caption{Examples of stationary chemoattractant distributions explored in Section~\ref{res:Euclid}. The setting on the left is referred to as the ``weakly trapping'' experiment, whereas the one on the right corresponds to the ``strongly trapping'' experiment.}
\label{fig1}
\end{center}
\end{figure}

We begin our exploration of how fractional Brownian motion shapes the dynamics of chemotactic cell migration by first considering the Euclidean setting, that is, the case of a flat substrate. As discussed in the introduction and supported by experimental evidence, we focus on fractional Brownian motion with Hurst index $H>1/2$, which corresponds to positively autocorrelated temporal noise \cite{Korabel2022,Makarava:2014}. Our results will be compared to the baseline case $H=1/2$, where Eq.~\eqref{SDE2} is driven by Brownian motion. 

The stationary distribution and mixing times for special cases of Eq.~\eqref{SDE2} are well understood when $H=1/2$ \cite{Durmus,Matus:2017}. For example, when $\chi(\,\cdot\,) \equiv 1$ and $\mathbf{\Sigma}(\,\cdot\,) \equiv \sqrt{2}\,\mathbf{I}$, the stationary distribution of Eq.~\eqref{SDE2} has density 
\[
\pi(\mathbf{x}) \propto \exp\bigl(f(\mathbf{x})\bigr),
\]
provided that the right-hand side is integrable. This illustrates how the inclusion of noise in Eq.~\eqref{SDE2} transforms a local maximization algorithm (gradient ascent) into a global one: sample paths of Eq.~\eqref{SDE2} asymptotically wander toward spatial regions that maximize $\pi(\mathbf{x})$ globally \cite{Durmus,Matus:2017}. This property is of particular relevance for noisy chemotactic motion, since eukaryotic cells must navigate fluctuating gradients and maintain robustness against spatial perturbations in order to effectively explore their environment. 

\begin{figure}[!t]
    \centering
    \includegraphics[width=0.9\textwidth]{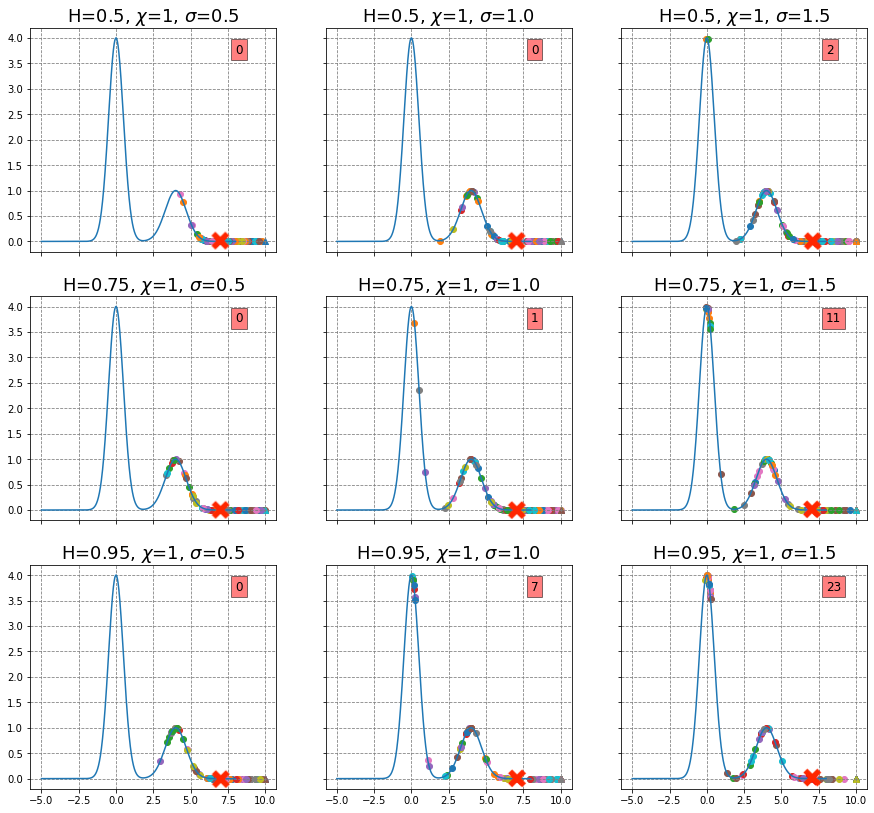}
    \caption{Spatial positions of $100$ particles at $t=10$ for different parameter sets in a one-dimensional version of the \emph{weakly} trapping experiment. An X-mark indicates the initial particle location, and the numbers in the red boxes denote the total number of particles that reached the global maximum within the observation window.}
    \label{fig2}
\end{figure}

In what follows, we investigate stationary chemoattractant profiles that model spatial fluctuations through the presence of local concentration maxima surrounding the signal origin, that is, the global maximum of the chemoattractant field, as exemplified in Fig.~\ref{fig1}. Our focus is on how positive temporal correlations in random exploratory behavior, and the associated persistence, accelerate the discovery of the global maximum by at least one cell in the population responding to the chemotactic cue. The underlying intuition is that stronger persistence, induced by temporal correlations, facilitates faster escape from local maxima. We refer to the distribution shown on the left of Fig.~\ref{fig1} as the ``weakly trapping'' experiment, and to the one on the right as the ``strongly trapping'' experiment.

Bell-shaped and spatially structured chemotactic profiles of the type considered in this section are not merely theoretical idealizations, but can also be generated experimentally in microfluidic settings. For example, \cite{Sawai:2016} studies \textit{Dictyostelium} chemotaxis in traveling waves of cAMP implemented as localized bell-shaped gradients. More generally, although many biological signaling fields are time-dependent, stationary or near-stationary gradients are routinely realized in controlled microfluidic chemotaxis experiments; see, for example, \cite{Amselem:2012,Asokan:2014}.

\begin{figure}[!t]
    \centering
    \includegraphics[width=0.97\textwidth]{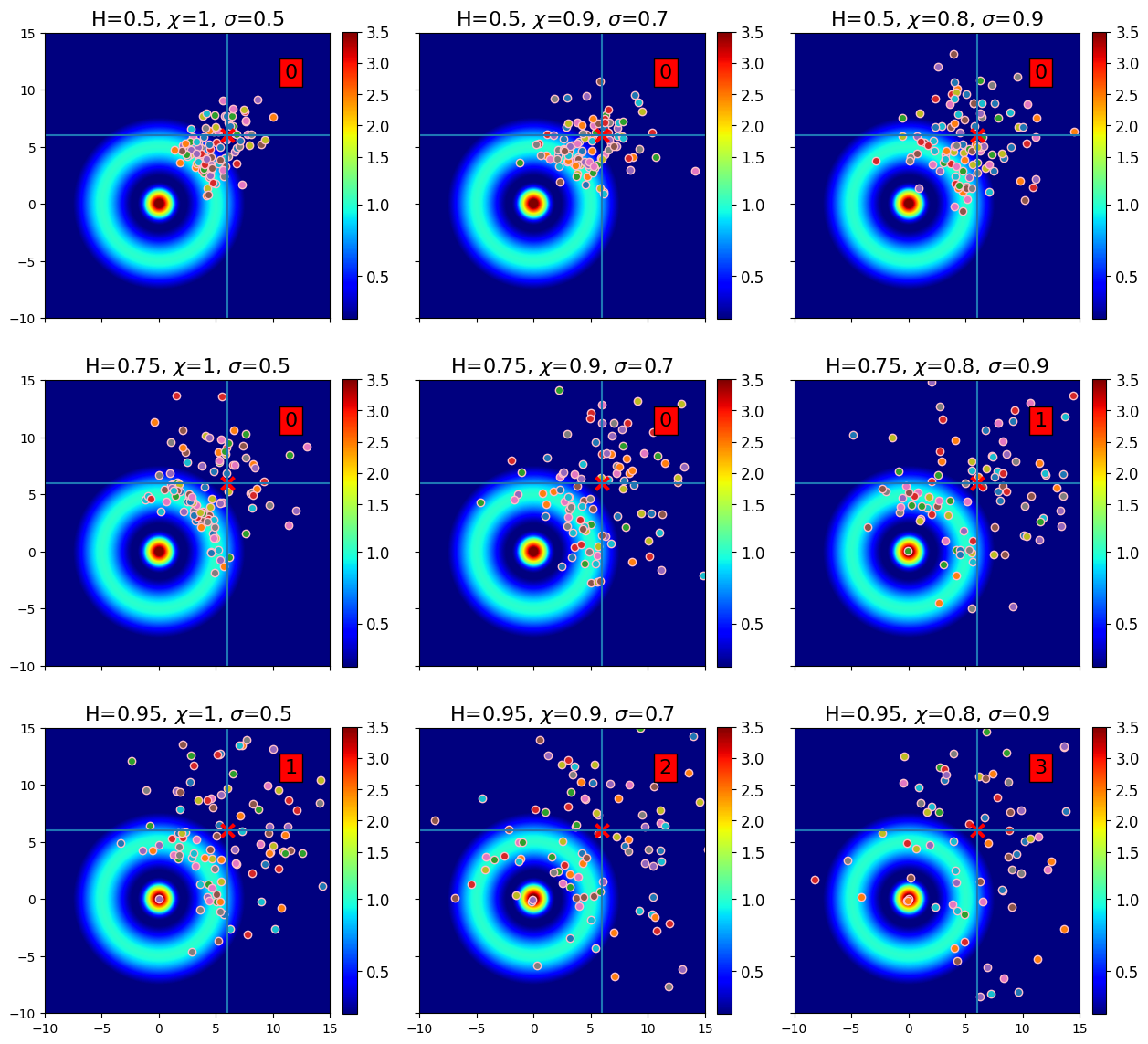}
    \caption{Spatial positions of $100$ particles at $t=10$ for different parameter sets in a two-dimensional version of the \emph{weakly} trapping experiment of Fig.~\ref{fig2}. An X-mark indicates the initial particle location, and the numbers in the red boxes denote the total number of particles that reached the global maximum within the observation window.}
    \label{fig3}
\end{figure}

Spatially structured and even non-monotone chemoattractant landscapes can also arise in heterogeneous biological tissues through localized secretion, receptor-mediated scavenging or absorption, extracellular-matrix binding, and tissue geometry. For instance, \cite{Spinosa:2017} showed, in a model of CXCL12/CXCR7 signaling in tumors, that the local composition of secreting and scavenging cells can control both the magnitude and the direction of CXCL12 gradients; \cite{Torisawa:2010} demonstrated experimentally that CXCL12-producing cells and CXCR7-expressing sink cells can generate source--sink chemokine gradients in microfluidic devices; and \cite{Boldajipour:2008} showed that CXCR7-mediated sequestration of SDF-1/CXCL12 in the zebrafish somatic environment is required to shape the guidance field for primordial germ-cell migration. Related work on VEGF transport further illustrates how realistic tissue geometry, extracellular-matrix binding, receptor binding, and internalization can produce spatially heterogeneous protein distributions and receptor activation patterns \cite{MacGabhann:2006}. Such heterogeneous source--sink landscapes need not guide cells monotonically toward a single target; rather, local maxima or competing gradients may transiently retain, redirect, or effectively ``trap'' chemotactic cells away from a desired region. Consistent with this interpretation, experiments with chemokine-releasing microspheres have shown that leukocytes can swarm around isolated local attractant sources for hours, and can hop between nearby sources when gradients overlap \cite{WangIrvine:2013}. In addition, there are biologically relevant settings in which the guidance field is effectively stationary over the timescale of the migration event under consideration, either because the signaling profile evolves sufficiently slowly relative to the migration dynamics or because the relevant cue is tissue-bound or otherwise spatially immobilized; see, for example, \cite{Schumann:2010,Schwarz:2017,Venkites,Weber:2013}. The analytic expressions for the stationary profiles considered in this section are collected in Appendix 3.

\begin{figure}[!t]
    \centering
    \includegraphics[width=0.9\textwidth]{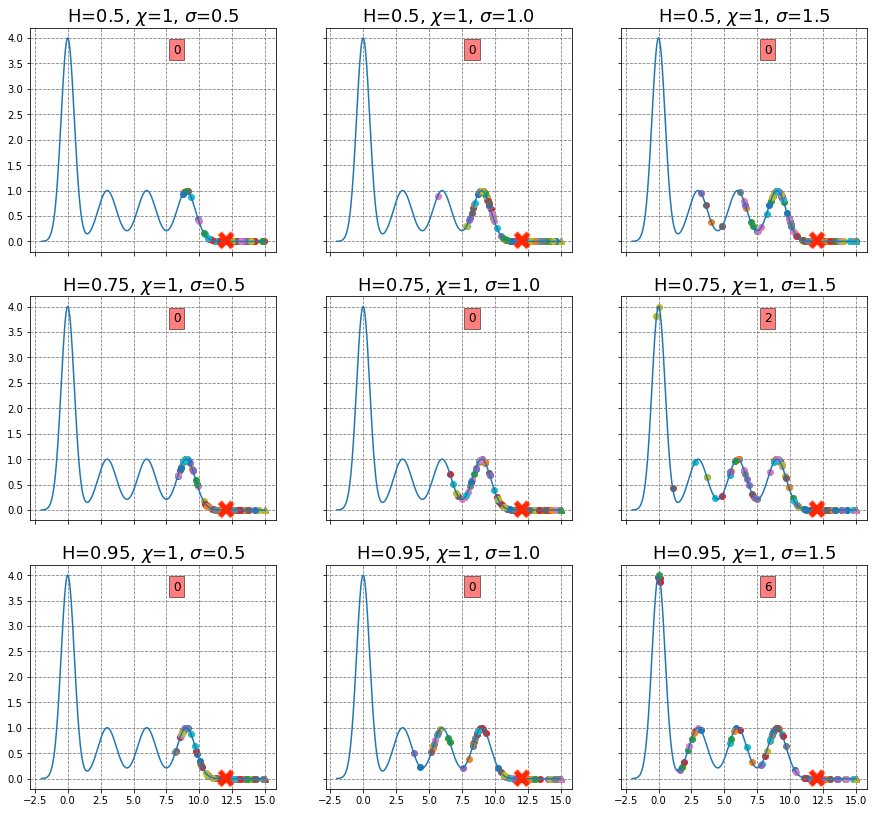}
    \caption{Spatial positions of $100$ particles at $t=10$ for different parameter sets in a one-dimensional version of the \emph{strongly} trapping experiment. An X-mark indicates the initial particle location, and the numbers in the red boxes denote the total number of particles that reached the global maximum within the observation window.}
    \label{fig4}
\end{figure}

Figures~\ref{fig2} and \ref{fig3} show the spatial positions of $100$ sample paths of Eq.~\eqref{SDE2} at time $t=10$ for different values of $H$, $\chi$, and $\sigma$ in one- and two-dimensional versions of the weakly trapping experiment, respectively. All numerical simulations were performed using an Euler-type scheme (see, e.g., \cite{Mishura_p2008}), with time step $\Delta t = 0.1$. The first raw in both figures corresponds to the setup, where Eq.~\eqref{SDE2} is driven by Brownian motion. The temporary ``trapping" of the sample paths in local maxima is especially apparent in Fig.   \ref{fig3}, where the trajectories located in the outer rim of maxima tend to move circumferentially, aligning with the convective dynamics of gradient ascent.  The parameters for chemotactic strengths and the noise amplitude constants for the experiments shown here were chosen to be slightly different in the one- and two-dimensional cases for visualization purposes, but similar results were obtained for a wide variety of parameter values around the ones shown here. 

In both the weakly and strongly trapping settings, we observe a consistent trend: as the Hurst index $H$ and the noise amplitude coefficient $\sigma$ increase, a larger number of particles reach the global maximum within the observation window. This behavior reflects the enhanced persistence imparted by positive temporal correlations and the stronger exploratory drive generated by increased stochastic fluctuations. The combination of these effects enables particles to escape local maxima more effectively and to converge toward the global maximum. 

\begin{figure}[!t]
    \centering
    \includegraphics[width=0.97\textwidth]{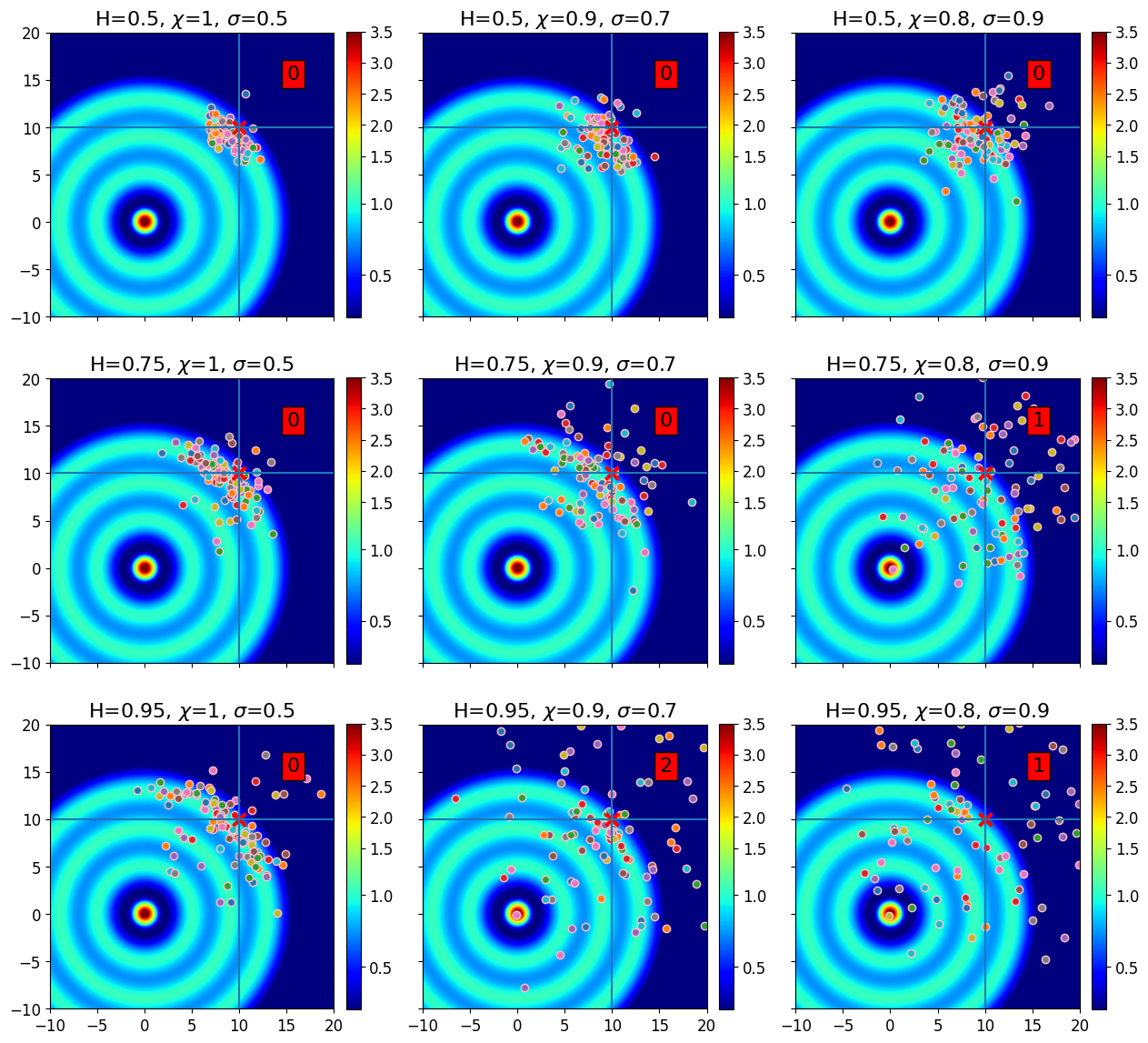}
    \caption{Spatial positions of $100$ particles at $t=10$ for different parameter sets in a two-dimensional version of the \emph{strongly} trapping experiment of Fig.~\ref{fig4}. An X-mark indicates the initial particle location, and the numbers in the red boxes denote the total number of particles that reached the global maximum within the observation window.}
    \label{fig5}
\end{figure}

We also note that some trajectories eventually leave the simulation domain. From both an optimization and a biological perspective, it is important to emphasize that success does not require every particle to locate the global maximum. In optimization, the discovery of the maximum by even a single trajectory suffices. In the biological setting, many cell lines exhibit collective strategies: once a cell reaches the global maximum and saturates its receptors with chemoattractant signaling molecules, it secretes a secondary chemoattractant that dynamically reshapes the chemical landscape and recruits additional cells to the same location \cite{Afonso2012,Devreotes1989,RE:2004}. This cooperative mechanism will be investigated in detail in Section~\ref{section:3.3}. Accordingly, all experiments reported in this section correspond to a randomly stopped version of Eq.~\eqref{SDE2}, in which simulated cells terminate their motility program once the chemoattractant concentration exceeds a threshold value $r>0$ ($r=3.5$ in the simulations shown). More precisely, the chemotactic sensitivity is taken to be $\chi\,\mathbbm{1}_{\{f(\mathbf{X}_t)\leq r\}}$ and the noise amplitude $\sigma\,\mathbbm{1}_{\{f(\mathbf{X}_t)\leq r\}}\,\mathbf{I}$, where $\chi>0$, $\sigma>0$, and $\mathbf{I}$ denotes the identity matrix. In this way, both the gradient-driven response and the random exploratory component are active only while the cell remains below the threshold, and are switched off once the cell reaches a sufficiently strong chemoattractant signal. This models the situation in which a cell localizes near the source after successful detection, rather than continuing to explore past it. As we explain in Section~\ref{section:3.3}, this mechanism is fundamentally different from the one considered there, where random motility remains active and localization instead arises through a secondary chemotactic cue.

\begin{figure}[!t]
\begin{center}
\hspace*{-0.3cm}\subfloat[][]{
    \includegraphics[width=0.49\textwidth]{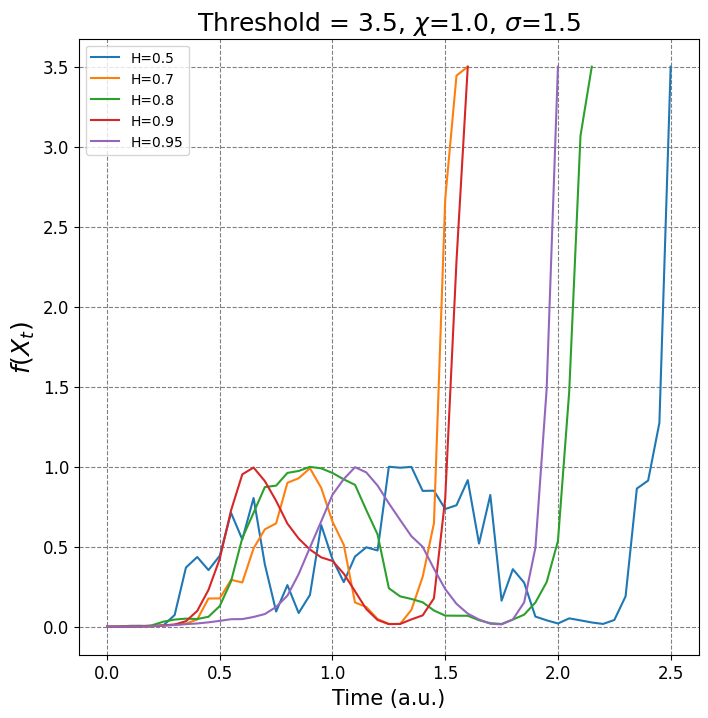}
}
\subfloat[][]{
    \includegraphics[width=0.49\textwidth]{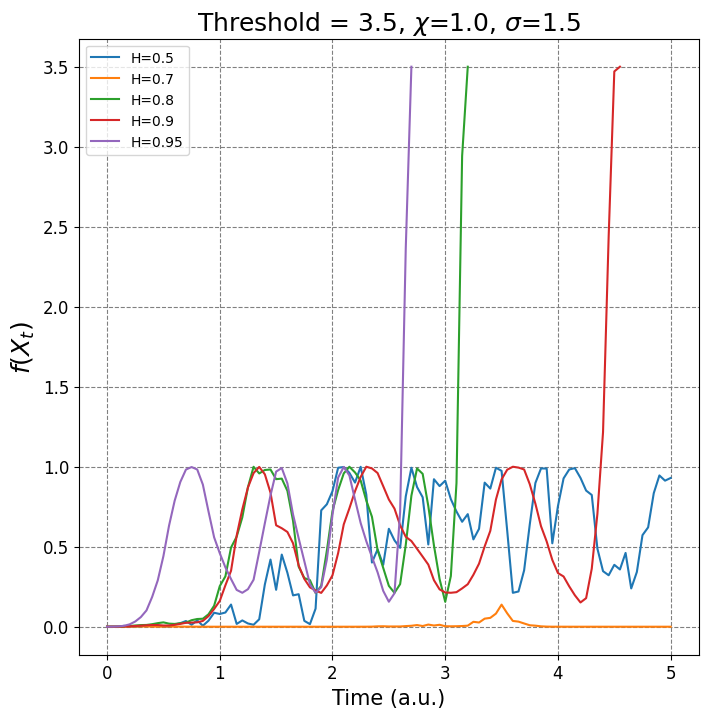}
}
\caption{Sample paths of $f(\mathbf{X}_t)$ in (A) the weakly trapping, one-dimensional experiment and 
(B) the strongly trapping, one-dimensional experiment. The $y$-axis represents the chemoattractant concentration sensed by individual cells at their location $\mathbf{X}_t$ at time $t$.}
\label{fig6}
\end{center}
\end{figure}

Figures~\ref{fig4} and \ref{fig5} show the corresponding results for the strongly trapping experiment, in both one and two dimensions. Compared to the weakly trapping case (Figs.~\ref{fig2} and \ref{fig3}), the additional local maxima present a more challenging environment, leading to more pronounced trapping of trajectories in secondary peaks. Nonetheless, the same overall trend persists: increasing positive temporal correlations and diffusion enhances the likelihood that at least one particle escapes local maxima and reaches the global maximum.

\begin{figure}[!t]
    \centering
    \hspace*{-0.8cm}\includegraphics[width=1.0\textwidth]{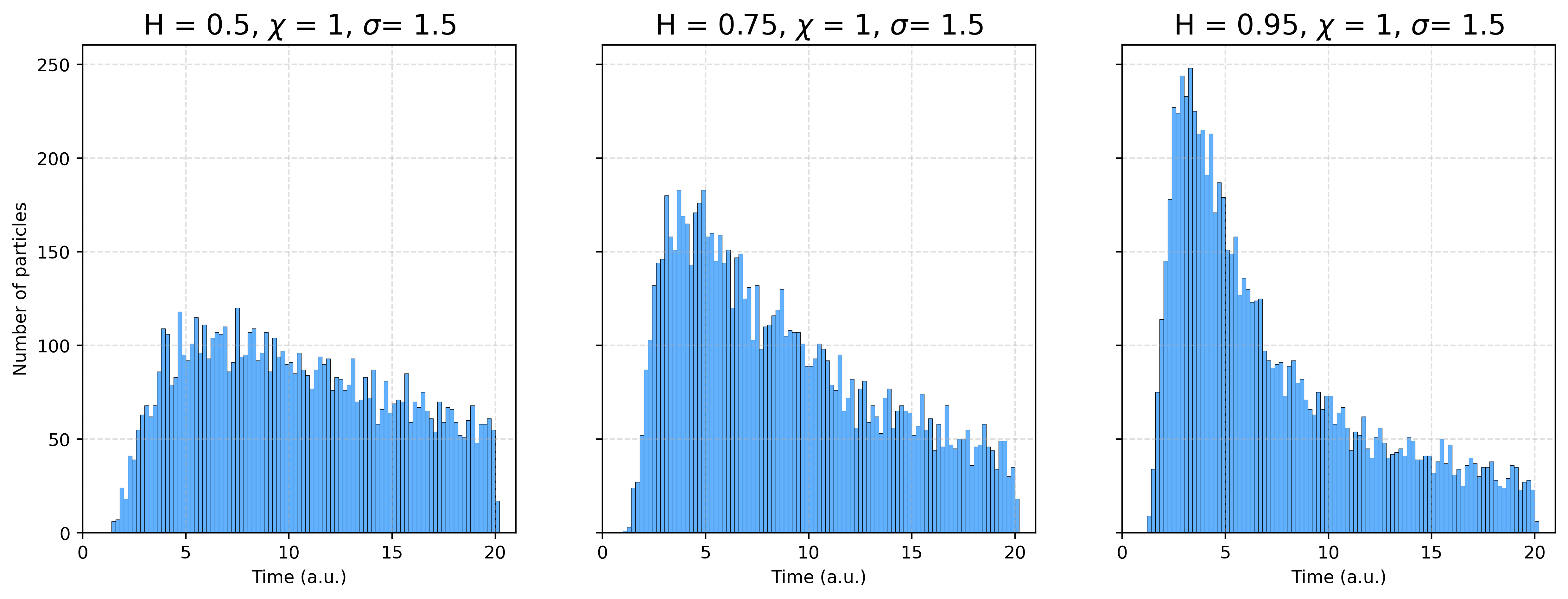}
    \caption{Unscaled histograms for the empirical distributions of the hitting time $T$ conditioned on the event $T\leq 20$ for different Hurst indices in the one-dimensional,  \emph{weakly} trapping experiment. The simulations were initialized with $20,000$ particles.}
    \label{fig7}
\end{figure}
\begin{figure}[!t]
    \centering
    \hspace*{-0.8cm}\includegraphics[width=1.0\textwidth]{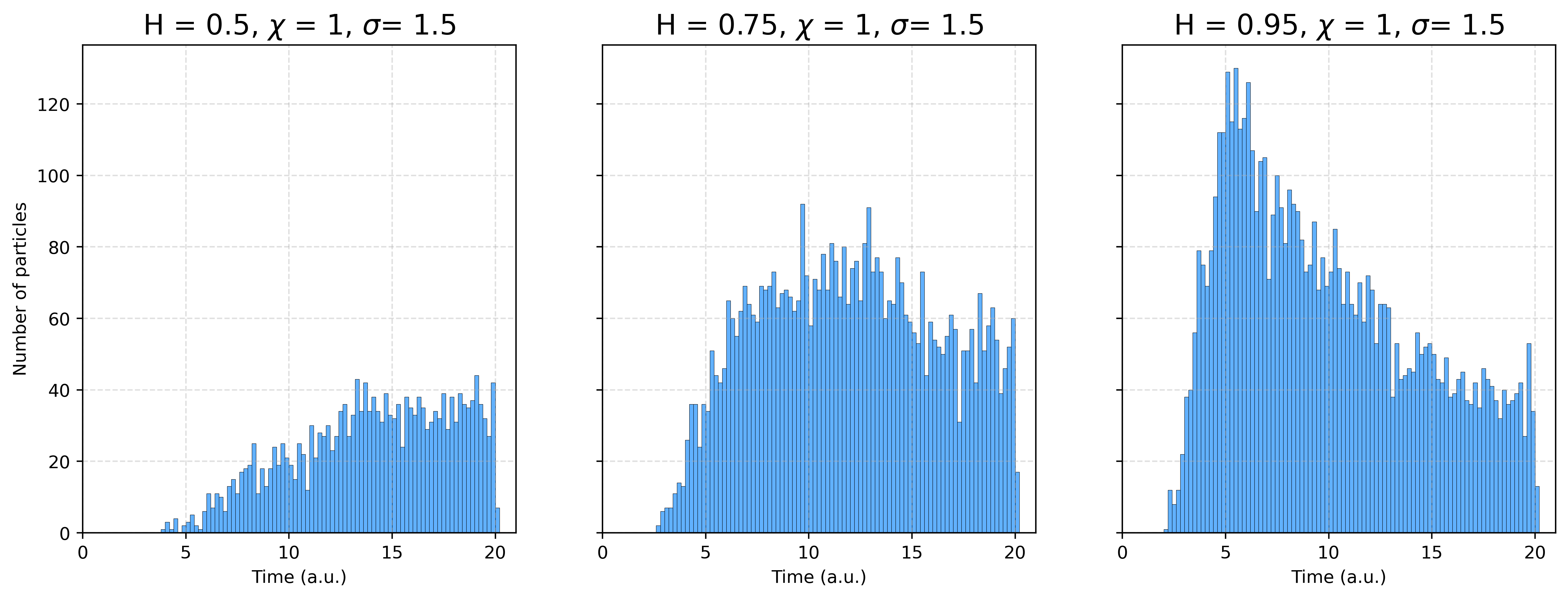}
    \caption{Unscaled histograms for the empirical distributions of the hitting time $T$ conditioned on the event $T\leq 20$ for different Hurst indices in the one-dimensional, \emph{strongly} trapping experiment. The simulations were initialized with $20,000$ particles.}
    \label{fig8}
\end{figure}

\begin{figure}[!t]
    \centering
    \hspace*{-0.8cm}\includegraphics[width=1.0\textwidth]{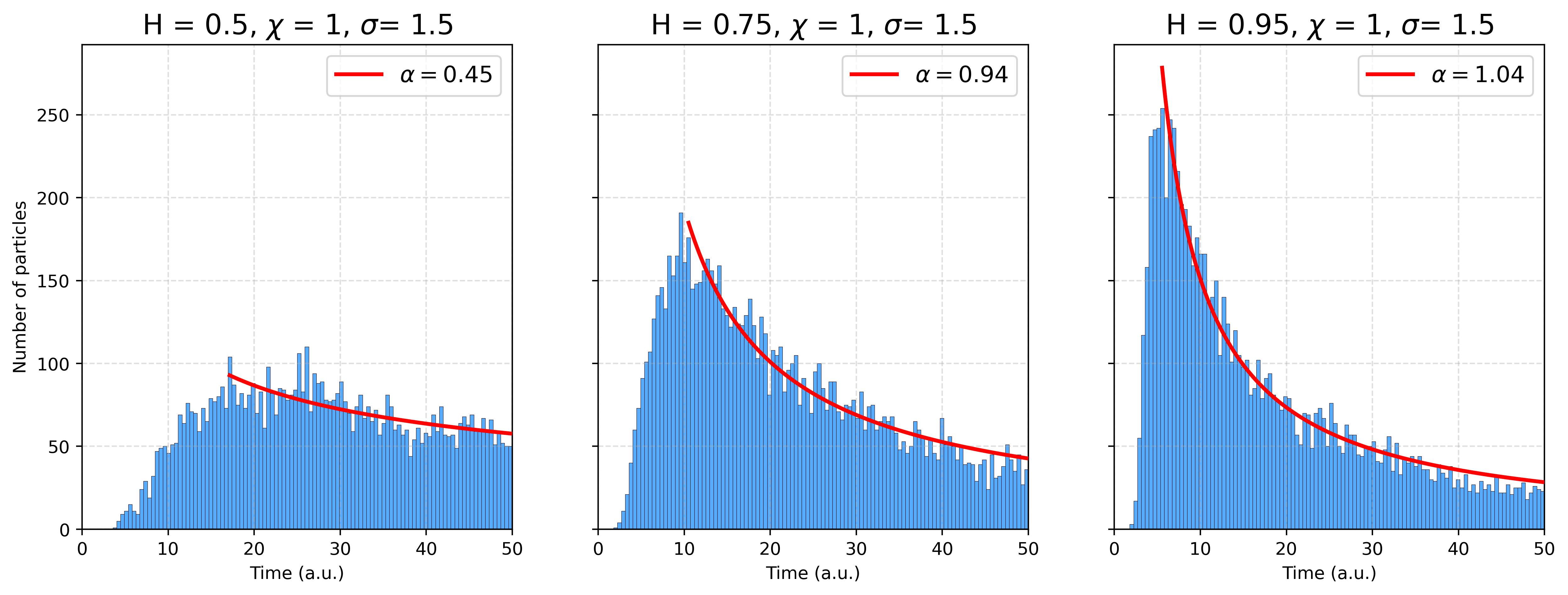}
    \caption{Unscaled histograms for the experiment shown in Fig. \ref{fig7} conditioned on the event $T\leq 50$.  The distributions, albeit quantitatively different over different observation windows, they remain qualitatively similar and their tails scale as a power law of the form $c (x-m)^{-\alpha}$. The different values of $\alpha$ for different Hurst indices are shown in the figure.}
    \label{fig9}
\end{figure}
\begin{figure}[!t]
    \centering
    \hspace*{-0.8cm}\includegraphics[width=0.7\textwidth]{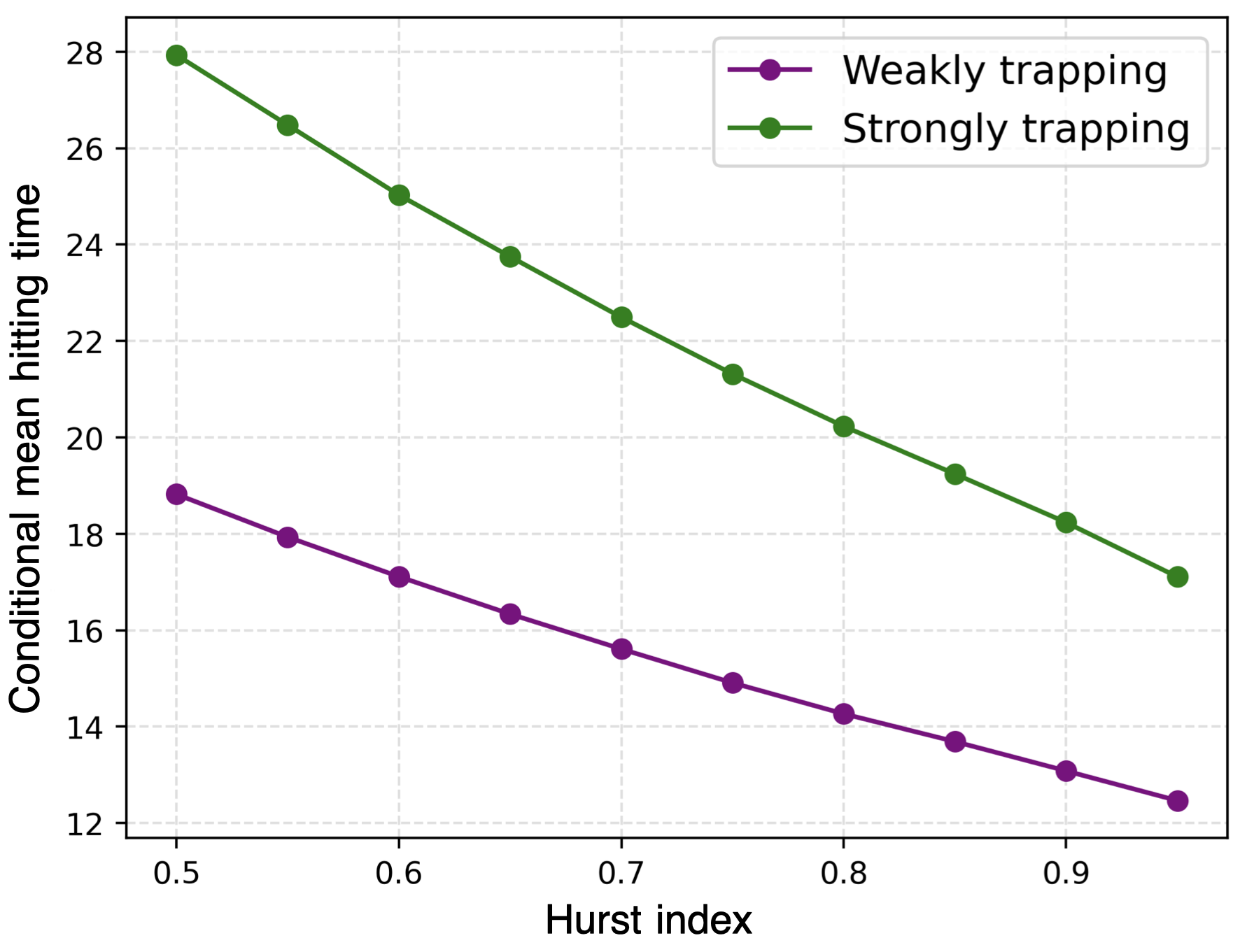}
     \caption{Dependence of the conditional mean hitting time (conditioned on $T\leq 50$) on the Hurst index $H>1/2$ for the weakly and strongly trapping experiments.}
    \label{fig10}
\end{figure}

To gain a more detailed understanding of how the Hurst index $H$ influences the dynamics of Eq.~\eqref{SDE2}, Fig.~\ref{fig6} shows selected sample paths of $f(\mathbf{X}_t)$ in the weakly and strongly trapping regimes for different values of $H$. Once again, it is apparent that Hurst indices greater than $1/2$ facilitate better exploration of the chemoattractant landscape by allowing for faster escape from local maxima. 

An interesting phenomenon can be observed in Fig.~\ref{fig6}(A), where increasing $H$ does not produce a monotonic decrease in the time required to reach the global maximum. The trajectories in orange and yellow ($H = 0.7$ and $H = 0.9$) arrive sooner than those in green and purple ($H = 0.8$ and $H = 0.95$). Although individual paths do not capture the full distribution of hitting times, the reasons for these delays are instructive. The purple trajectory ($H = 0.95$) initially persists in moving away from the chemoattractant source, postponing detection of the gradient. The green trajectory ($H = 0.8$), by contrast, advances in the correct direction but with increments that are atypically small during the interval from about $t = 0.6$ to $t = 1.2$; because of temporal correlations, these reduced step sizes persist throughout that period, slowing progress toward the maximum. These cases illustrate that while positive correlations typically aid escape from local maxima, they can also delay arrival either by reinforcing motion away from the chemoattractant source or by sustaining unusually small advances along the correct direction for a finite period of time.

\begin{figure}[!t]
    \centering
    \includegraphics[width=0.98\textwidth]{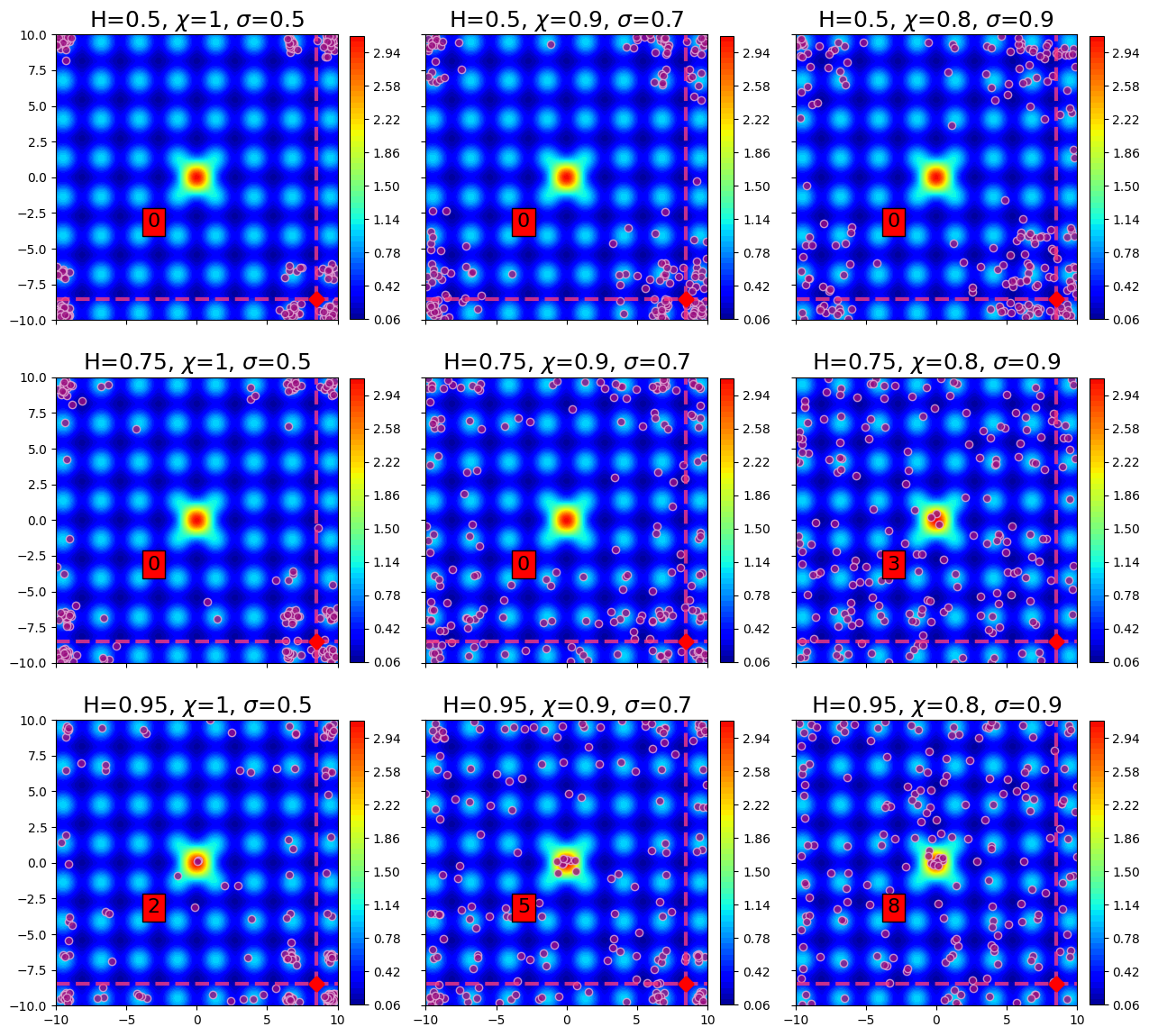}
    \caption{Spatial positions of $200$ particles at $t=15$ for different parameter sets in a two-dimensional, multimodal chemotactic field. The intersection of the dashed red lines marks the initial particle location. Periodic boundary conditions were imposed. The numbers in the red boxes indicate the total number of particles that reached the global maximum during the observation window.}
    \label{fig11}
\end{figure}

In order to quantify the phenomena observed in this section and to capture the collective dynamics, we computed the empirical distributions of the hitting time 
$$T = \inf \bigl\{ t \geq 0 : f(\mathbf{X}_t)\geq r \bigr\}$$
for different Hurst indices, where $r=3.5$ is the threshold above which the cells terminate their chemotactic response. Figures~\ref{fig7} and \ref{fig8} show the resulting distributions, conditioned on the event $T\leq 20$, from simulations initialized with $20,000$ particles in the weakly and strongly trapping experiments, respectively. In both settings, increasing the Hurst index leads to a greater number of particles reaching the global maximum within the observation window $[0,20]$. As $H$ increases, the distributions taper off toward the right tail: while a fraction of particles persist in moving away from the global maximum due to long-range correlations, a larger proportion succeed in reaching the maximum more rapidly. These findings complement the qualitative observations in Figs.~\ref{fig2}–\ref{fig6} and will be compared in Section~\ref{section:3.3} to experiments that incorporate collaborative dynamics through a secondary chemotactic signal.

Since these distributions are conditioned on successful arrival within a fixed observation window, it is natural to ask how their shape changes as that window is varied. To further assess the influence of the observation window, that is, the conditioning of the hitting-time distribution on an event of the form $T\leq T_{\text{obs}}$, we repeated the computation of the conditional hitting-time distributions for larger values of $T_{\text{obs}}$. Although the resulting distributions differ quantitatively from those shown in Figs.~\ref{fig7} and \ref{fig8}, as expected, they display the same qualitative behavior and in particular exhibit power-law decay in their tails. Figure~\ref{fig9} shows the corresponding (unscaled) conditional hitting-time distributions for the weakly trapping experiment when $T_{\text{obs}}=50$. In order to quantify the tail behavior, we fitted the distributions to shifted power laws of the form $c(x-m)^{-\alpha}$ for different Hurst indices. Across all numerical experiments, the fitted exponent $\alpha$ increased monotonically with $H$ for fixed $T_{\text{obs}}$, indicating that the conditional hitting-time distributions become lighter-tailed as the temporal correlations become stronger.

A complementary summary is provided in Fig.~\ref{fig10}, which shows the dependence of the conditional mean hitting time on the Hurst index for both the weakly and strongly trapping experiments when conditioning on the event $T\leq 50$. In both cases, the mean hitting time decreases monotonically as $H$ increases, confirming that stronger persistence leads not only to a larger number of successful trajectories within the observation window, but also to earlier arrivals on average among those trajectories. The same qualitative trend was observed for other parameter values and for other observation windows. Taken together, these observations show that the beneficial effect of increasing $H$ is robust with respect to the choice of observation window, both at the level of the tail behavior of the conditional hitting-time distributions and at the level of their conditional means.

It is useful to compare these results with classical one-dimensional first-passage results for free fractional Brownian motion. In that setting, the unconditional hitting-time distribution has a heavy long-time tail, with persistence becoming stronger for $H>1/2$ than in the Brownian case \cite{Ding:1995,Molchan:1999}. This observation does not contradict the conclusions of the present paper, because the quantity analyzed here is different: our simulations report the empirical distribution of the hitting time $T$ conditioned on the event $T\leq T_{\mathrm{obs}}$, where $T_{\mathrm{obs}}$ is a finite observation window. The numerical results shown correspond to $T_{\mathrm{obs}}=20$ and $T_{\mathrm{obs}}=50$, and the same qualitative behavior was observed for larger windows up to $T_{\mathrm{obs}}=1000$ (not shown). Conditional first-passage statistics of this type are also well established in Markovian settings. For example, \cite{Lawley:2021} studied mortal diffusive searchers and showed that conditioning on successful target detection before an inactivation time can make the conditional first-passage time substantially faster than the corresponding first-passage time without inactivation  \cite{Lawley:2021,Isaacson:2020}.

The distinction can be understood heuristically as follows. For persistent motion, trajectories that initially move toward the target tend to maintain this favorable direction and therefore arrive sooner than in the Brownian case, whereas trajectories that initially move away may undergo long excursions before returning. The latter trajectories contribute to the heavy tail of the unconditional first-passage distribution, while the former can dominate the conditional distributions considered here. To provide a minimal baseline for this effect, Appendix 4 analyzes the one-dimensional problem and identifies the scaling regime in which the hitting time of a target by fractional Brownian motion with $H>1/2$, conditioned on detection before a finite observation time, is faster than the corresponding Brownian conditional hitting time. In the full simulations considered here, the particles also evolve in a chemotactic landscape rather than in free space, so the chemoattractant field provides directional information and persistence can further facilitate escape from local trapping regions.

To further assess the robustness conferred by positive temporal correlations against spatial fluctuations in the chemotactic landscape, we examined additional configurations of the chemoattractant field, including the multimodal distribution shown in Fig.~\ref{fig11}. The behavior observed in the weakly and strongly trapping settings persists in this more complex environment. Entrapment by local maxima is evident, particularly for lower values of $H$ and $\sigma$, yet increasing the Hurst index enhances the ability of particles to escape these regions and reach the global maximum of the chemoattractant concentration within the observation window. The empirical hitting-time distributions (not shown here) exhibit trends consistent with those reported above, confirming that higher values of $H$ systematically increase the proportion of successful trajectories and reinforce the overall robustness of the search dynamics.

These findings, together with the previous experiments, show that although higher Hurst indices can sustain unusually small advances along the correct direction for finite intervals of time, the dominant effect is an accelerated escape from local fluctuations, which increases the likelihood that particles discover the global maximum earlier. Building on these observations, we now turn to curved substrates, where geometry introduces additional directional biases into the dynamics.

\begin{figure}[!t]
    \centering
    \includegraphics[width=0.97\textwidth]{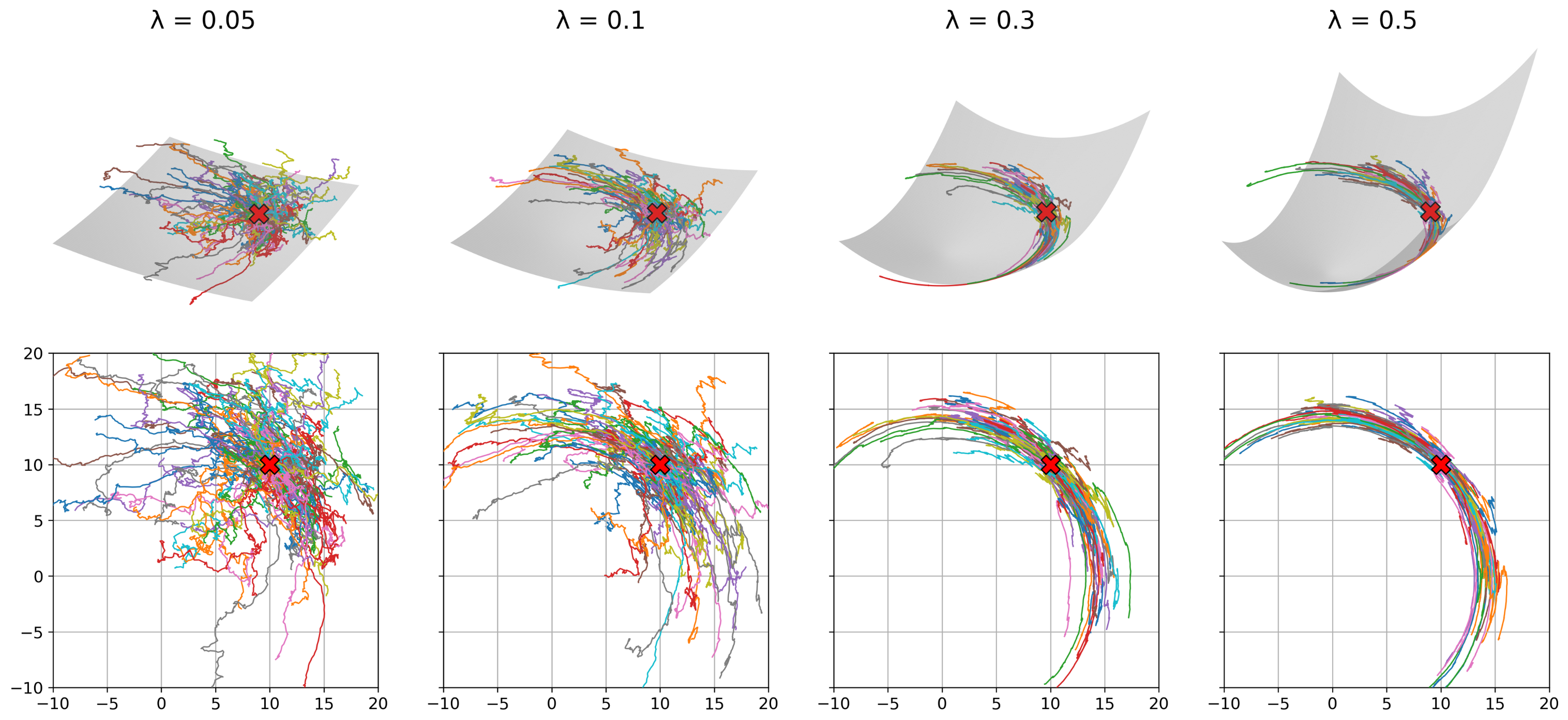}
    \caption{Sample paths of fractional Brownian motion with Hurst index $H=0.8$ on the paraboloid $\psi(x,y)=\lambda(x^2+y^2)$ for different values of $\lambda$. The top row shows the process on the manifold, while the bottom row displays the corresponding process in local coordinates.}
    \label{fig12}
\end{figure}

\subsection{Chemotactic dynamics on curved substrates}\label{res:man}

Recent studies have demonstrated that substrate curvature can strongly influence cell motility, giving rise to a behavior known as \textit{curvotaxis} \cite{Pieuchot:2018,Sadhu:2024}. He and Jiang \cite{He:2017} employed a three-dimensional mechanical model to show that cells migrate more persistently along concave surfaces and display reduced motility on convex ones. This curvature-dependent migration was further validated experimentally by Pieuchot \textit{et al.} \cite{Pieuchot:2018}, who observed that cells preferentially move toward concave valleys and avoid convex ridges, a process mediated by the coupling between the cytoskeleton and the nucleus. 

Complementing these findings, a minimal theoretical model based on vesicle dynamics with curvature-sensing proteins reproduced curvotactic behavior across sinusoidal and tubular geometries, suggesting that such motility patterns can emerge from basic physical interactions \cite{Sadhu:2024}. Additional studies combining experimental observations with simulations further confirmed that curvature, together with confinement and surface topography, modulates cell morphology, migration speed, and actin organization, highlighting the intricate interplay between geometric cues and cytoskeletal dynamics \cite{Winkler:2019}.

These findings naturally raise the question of whether the robustness to signal fluctuations observed in the Euclidean setting of Section~\ref{res:Euclid} persists in the presence of directional biases introduced by substrate curvature. While we do not explicitly model biophysical mechanisms such as cytoskeletal-nuclear coupling or curvature-sensing proteins in this paper, it is well established that substrate geometry influences the dynamics of Equations \eqref{SDE_m}, as outlined in Section~\ref{manifold}. In this section, we examine how curvature-induced dynamics interact with chemotactic gradients and temporally correlated noise. Our results provide further evidence that fractional Brownian motion confers robustness even in the presence of competing directional influences arising from substrate geometry.

\begin{figure}[t]
    \centering
    \includegraphics[width=0.97\textwidth]{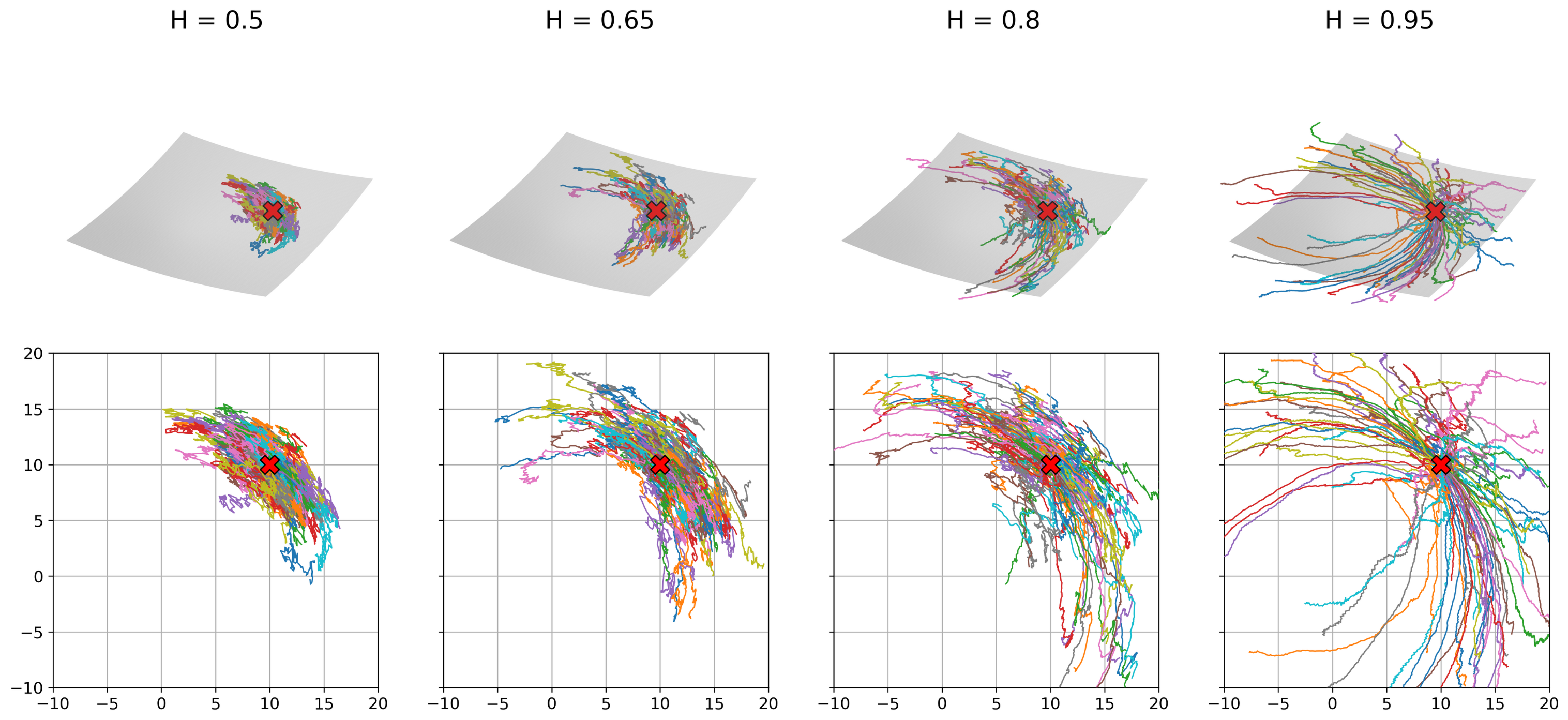}
    \caption{Sample paths of fractional Brownian motion for different values of the Hurst index on the paraboloid $\psi(x,y)=\lambda(x^2+y^2)$ with $\lambda = 0.1$. The top row shows the process on the manifold, while the bottom row displays the corresponding process in local coordinates.}
    \label{fig13}
\end{figure}

\begin{figure}[!t]
    \centering
    \includegraphics[width=0.65\textwidth]{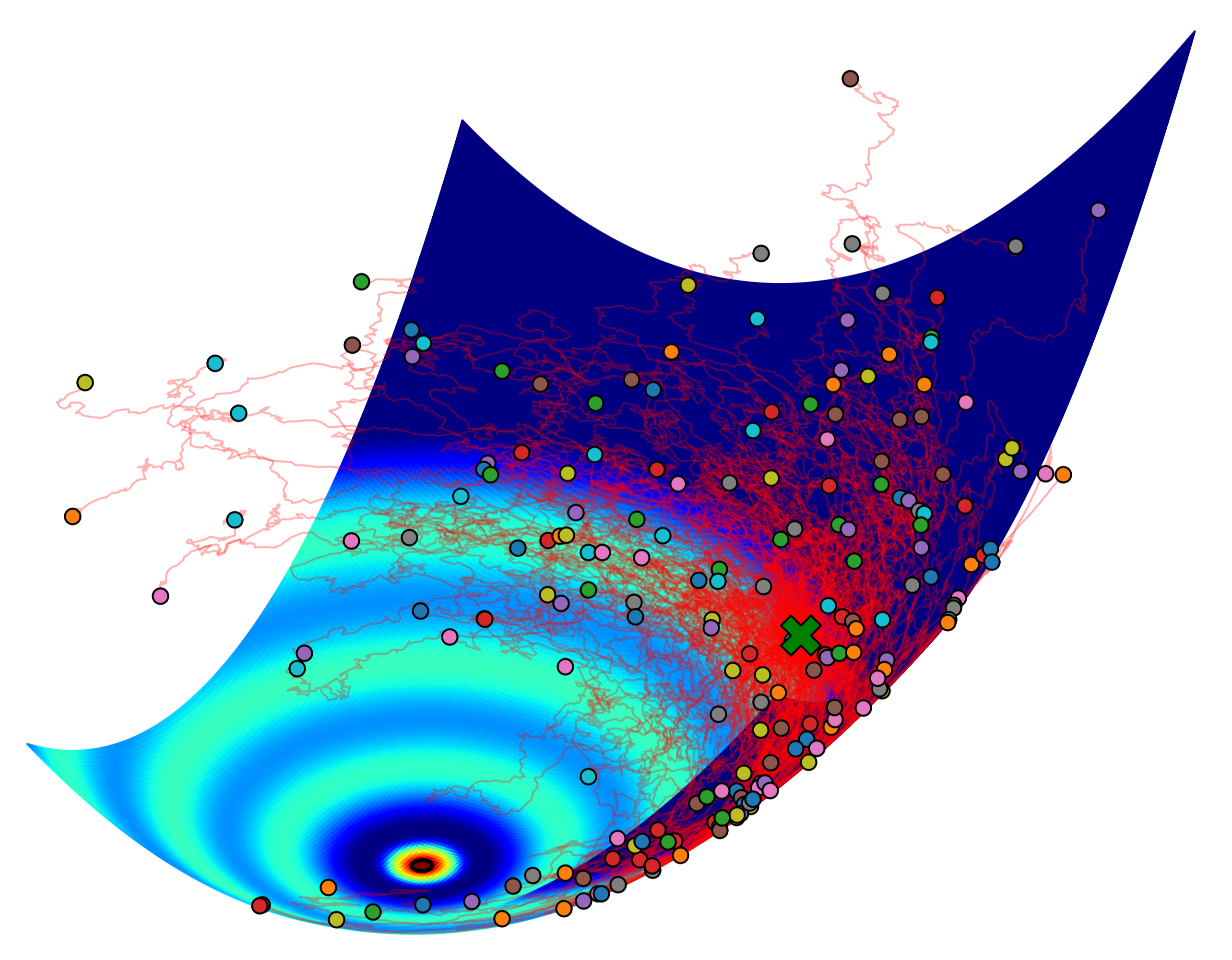}
    \caption{Sample paths of the solution to Eq.~\eqref{SDE_m} in the presence of a chemoattractant distribution on the manifold given by the graph of $\psi(x,y)=\lambda(x^2+y^2)$ with $\lambda=0.1$. An X-mark indicates the initial particle location.}
    \label{fig14}
\end{figure}

To illustrate how geometry influences fractional Brownian motion, Figs.~\ref{fig12} and \ref{fig13} present sample paths on paraboloids of varying curvatures and Hurst indices. The top rows depict trajectories on the manifolds, while the bottom rows show the corresponding processes in local coordinates, which in this case reduce to the plane on which the paraboloid is defined (see Section~\ref{manifold} and Appendix 1 for further details). The emergence of spatial correlations is clearly visible. For a fixed Hurst index $H \geq 1/2$, increasing the parameter $\lambda$ leads the process to align more strongly with the level sets of the paraboloid, producing predominantly circumferential motion. This behavior can be understood as a direct consequence of the surface geometry: as the particle moves farther from the origin, the tangent plane to the paraboloid becomes progressively steeper, so displacements orthogonal to the level sets diminish in relative importance compared to tangential ones. This effect illustrates how geometry can hinder chemotactic motion toward the global maximum of a chemoattractant distribution on the curved substrate (cf. Fig.~\ref{fig14}). At the same time, Fig.~\ref{fig13} shows that increasing the persistence of temporal correlations, achieved by raising the Hurst index, enables trajectories to overcome this geometric constraint and explore the manifold more broadly, thereby mitigating the hindering effect of curvature.

\begin{figure}[!t]
    \centering
    \includegraphics[width=0.97\textwidth]{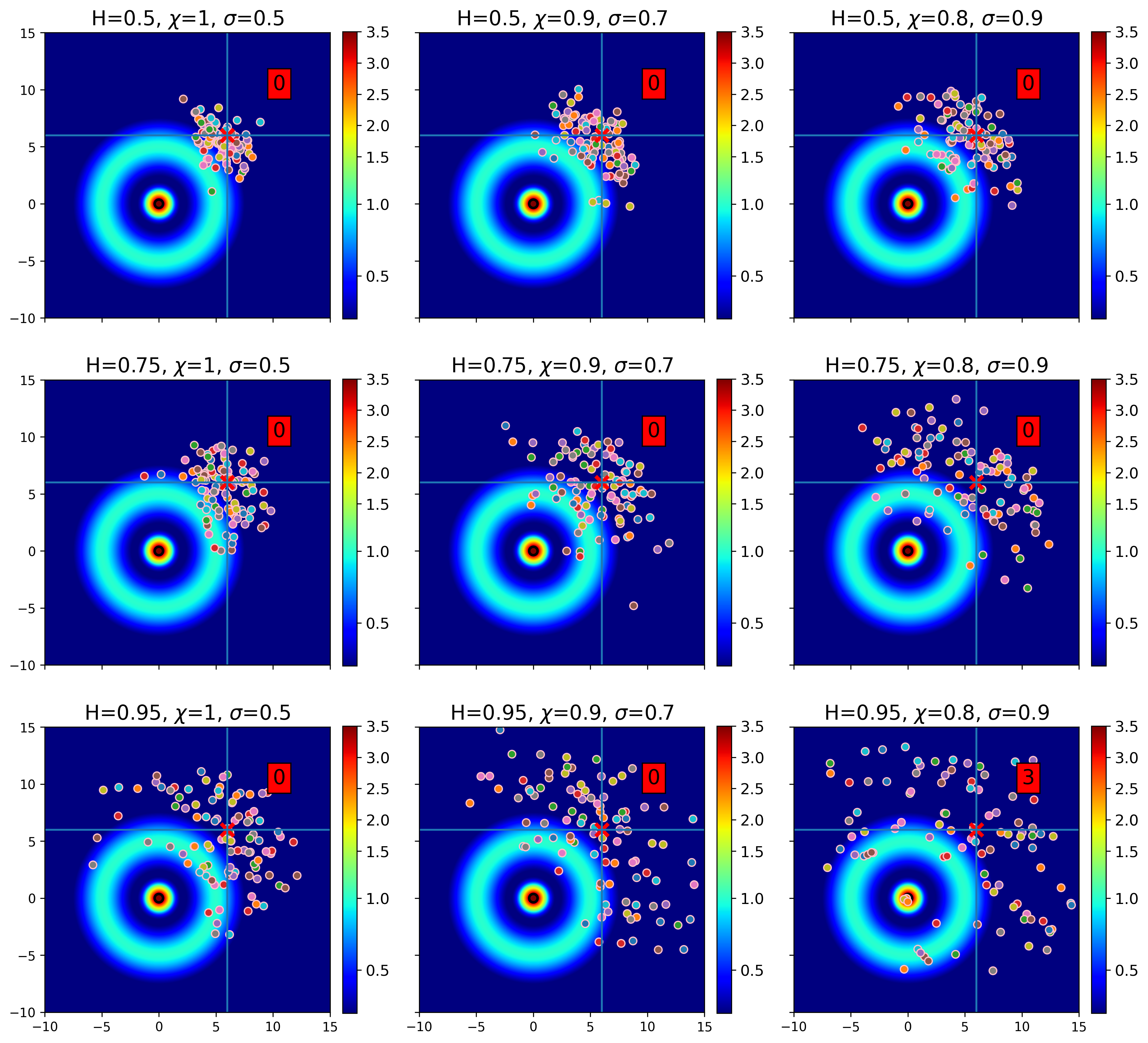}
    \caption{Spatial positions of $100$ particles on the paraboloid defined in the text (displayed in local coordinates) at $t=10$ for different parameter sets in the weakly trapping regime. An X-mark indicates the initial particle location, and the numbers in the red boxes denote the total number of particles that reached the global maximum within the observation window.}
    \label{fig15}
\end{figure}

\begin{figure}[!t]
    \centering
    \includegraphics[width=0.97\textwidth]{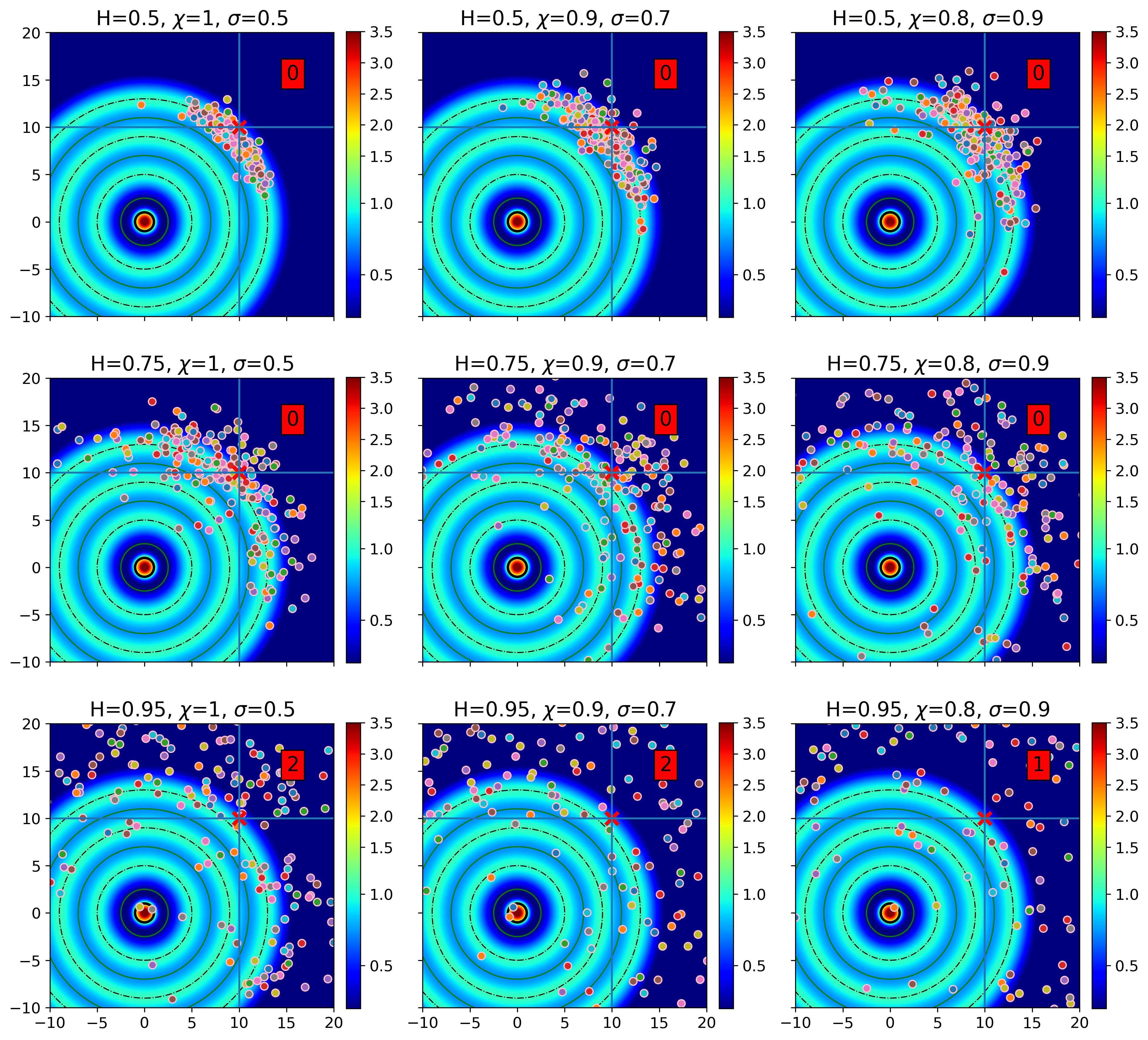}
    \caption{Spatial positions of $200$ particles on the paraboloid defined in the text (displayed in local coordinates) at $t=40$ for different parameter sets in the strongly trapping regime. An X-mark indicates the initial particle location, and the numbers in the red boxes denote the total number of particles that reached the global maximum within the observation window.}
    \label{fig16}
\end{figure}

To investigate the dynamics arising from the combined effects of geometry, fractional Brownian motion, and gradient ascent, we now revisit the two-dimensional experiments of Section~\ref{res:Euclid}. The weakly trapping and strongly trapping settings shown in Figs.~\ref{fig3} and \ref{fig5}, respectively, are repeated here on the paraboloid $\psi(x,y)=\lambda(x^2+y^2)$ with $\lambda=0.1$, as illustrated in Fig.~\ref{fig14}. The stationary chemoattractant distributions are the same as those used in Section~\ref{res:Euclid}, but here they are defined in local coordinates (i.e., on the plane) and subsequently projected onto the manifold. The realizations of chemotactic motion driven by Eq.~\eqref{SDE_m} in Figs.~\ref{fig15} and \ref{fig16} were obtained by solving Eq.~\eqref{SDE_m} in local coordinates, using the same numerical scheme and time step ($\Delta t=0.1$) as in Section~\ref{res:Euclid}. 

For reference, the metric tensor of the paraboloid $\psi$, viewed as an embedded manifold, is given by (see, e.g., Section 4.2.6 in \cite{Chow2})
$$
g_{ij} = \delta_{ij} + \partial_i \psi \, \partial_j \psi,
$$
where the usual tensor notation is used, and the indices $i$ and $j$ range over $1$ and $2$, corresponding to the $x$ and $y$ local coordinates. The inverse metric then takes the form
$$
g^{ij} = \delta^{ij} - \frac{\partial^i \psi \, \partial^j \psi}{\det(g)},
$$
where $\partial^i \psi := \sum_j \delta^{ij}\partial_j \psi$. This in turn implies that
\begin{equation}\label{sqinvm}
\left(g^{-1/2}\right)^i_j = \delta^i_j - \frac{\partial^i \psi \, \partial_j \psi}{\det(g) + \sqrt{\det(g)}},
\end{equation}
with a detailed derivation discussed in Appendix 2. Finally, the Christoffel symbols of the paraboloid, which enter the dynamics only in the case of Brownian motion, i.e., when $H=1/2$, are given by
$$
\Gamma_{ij}^k = \frac{\partial^k \psi \, \partial_i \partial_j \psi}{\det(g)}
$$

Figure~\ref{fig15} shows the spatial positions of $100$ sample paths of Eq.~\eqref{SDE_m} at time $t=10$ for different values of $H$, $\chi$, and $\sigma$ in the weakly trapping regime. Compared to the Euclidean case in Fig.~\ref{fig3}, the influence of geometry is apparent: trajectories exhibit circumferential motion that limits their ability to explore the region containing the global maximum. Trapping in secondary, ``outer'' local maxima is also observed, consistent with the Euclidean setting. However, the third row of Fig.~\ref{fig15} demonstrates that strong temporal correlations associated with high values of the Hurst index enable trajectories to overcome both the geometric hindrance and the local maxima of the chemotactic distribution, thereby accelerating convergence to the global maximum.

Figure~\ref{fig16} focuses on the strongly trapping regime, where sample paths must escape from several local maxima before reaching the global maximum at the origin. In this setting, the spatial correlations imposed by the geometry made the number of particles and time horizon used in the Euclidean experiments insufficient for capturing a trajectory that reached the global maximum. For this reason, the experiment was conducted with $200$ realizations over the time interval $[0,40]$. Both the circumferential bias induced by the geometry and the trapping in local maxima are clearly visible. Nevertheless, the final row of Fig.~\ref{fig16} shows that high values of the Hurst index confer robustness against both signal fluctuations, represented by the local maxima, and geometric constraints. In this case, several trajectories with large Hurst index successfully reach the global maximum within the observation window.

\begin{figure}[!t]
    \subfloat[][]{
    \includegraphics[width=0.34\textwidth]{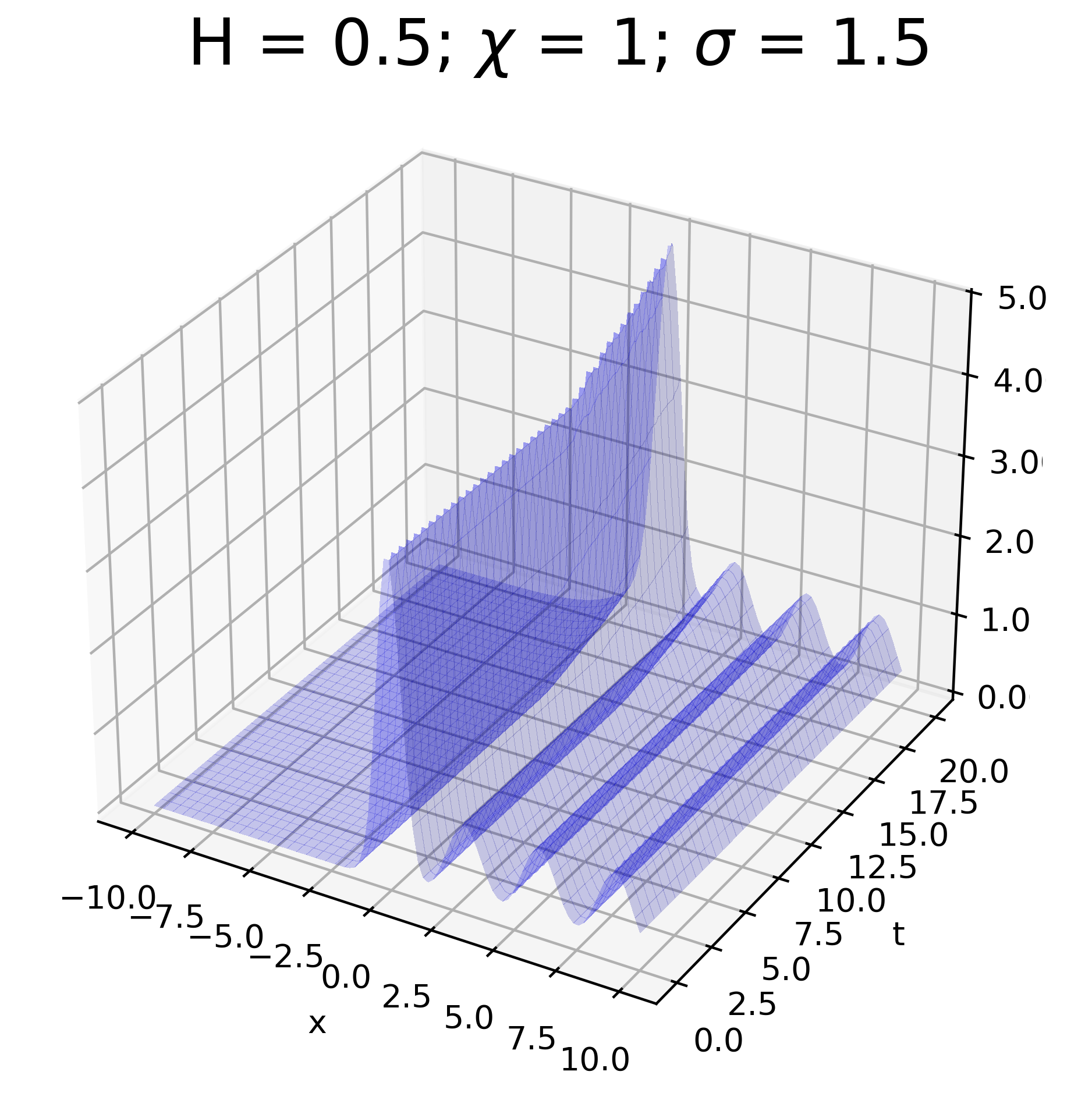}}
    \hspace*{-0.15cm}\subfloat[][]{
    \includegraphics[width=0.34\textwidth]{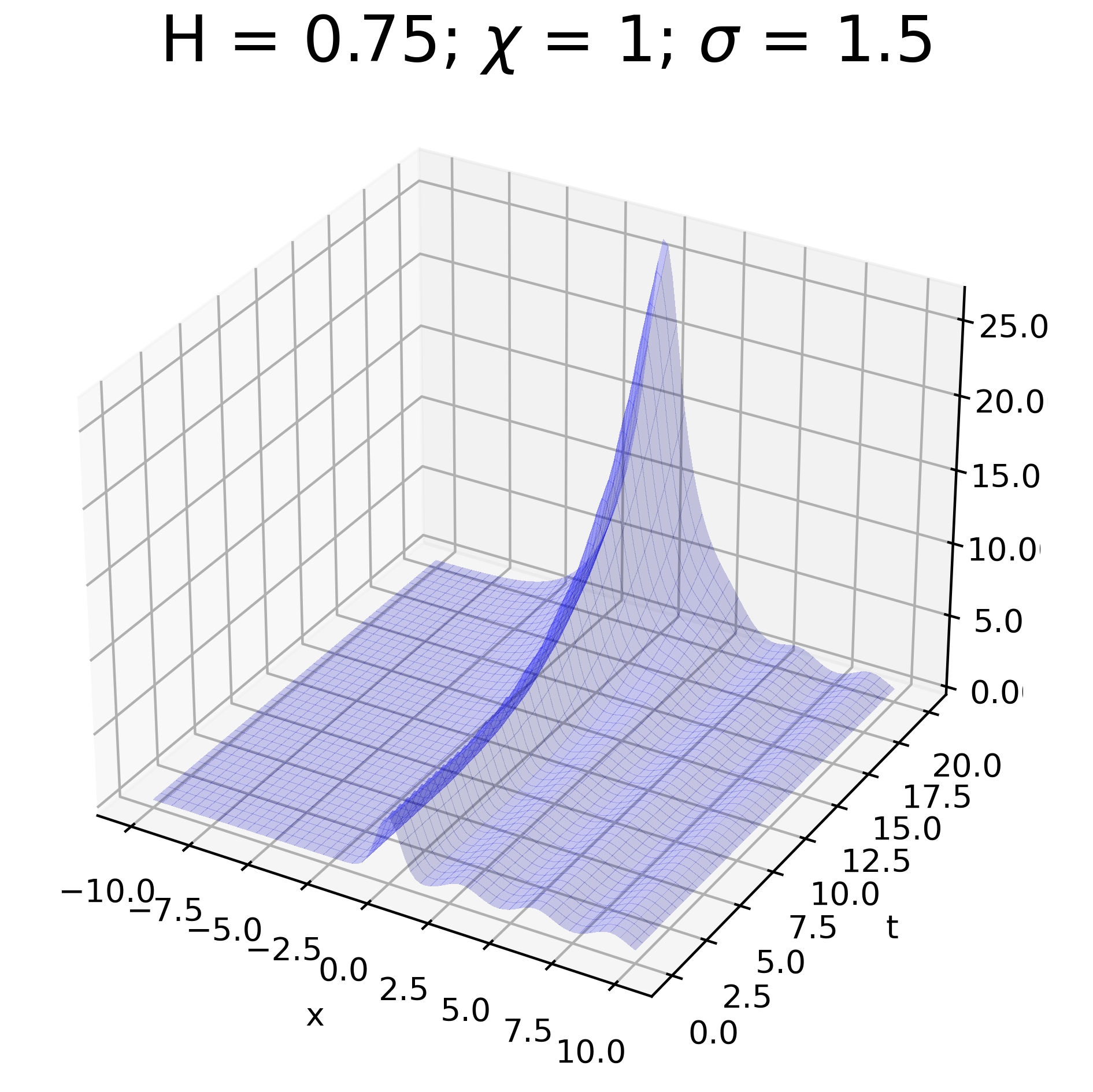}}
    \hspace*{-0.15cm}\subfloat[][]{
    \includegraphics[width=0.34\textwidth]{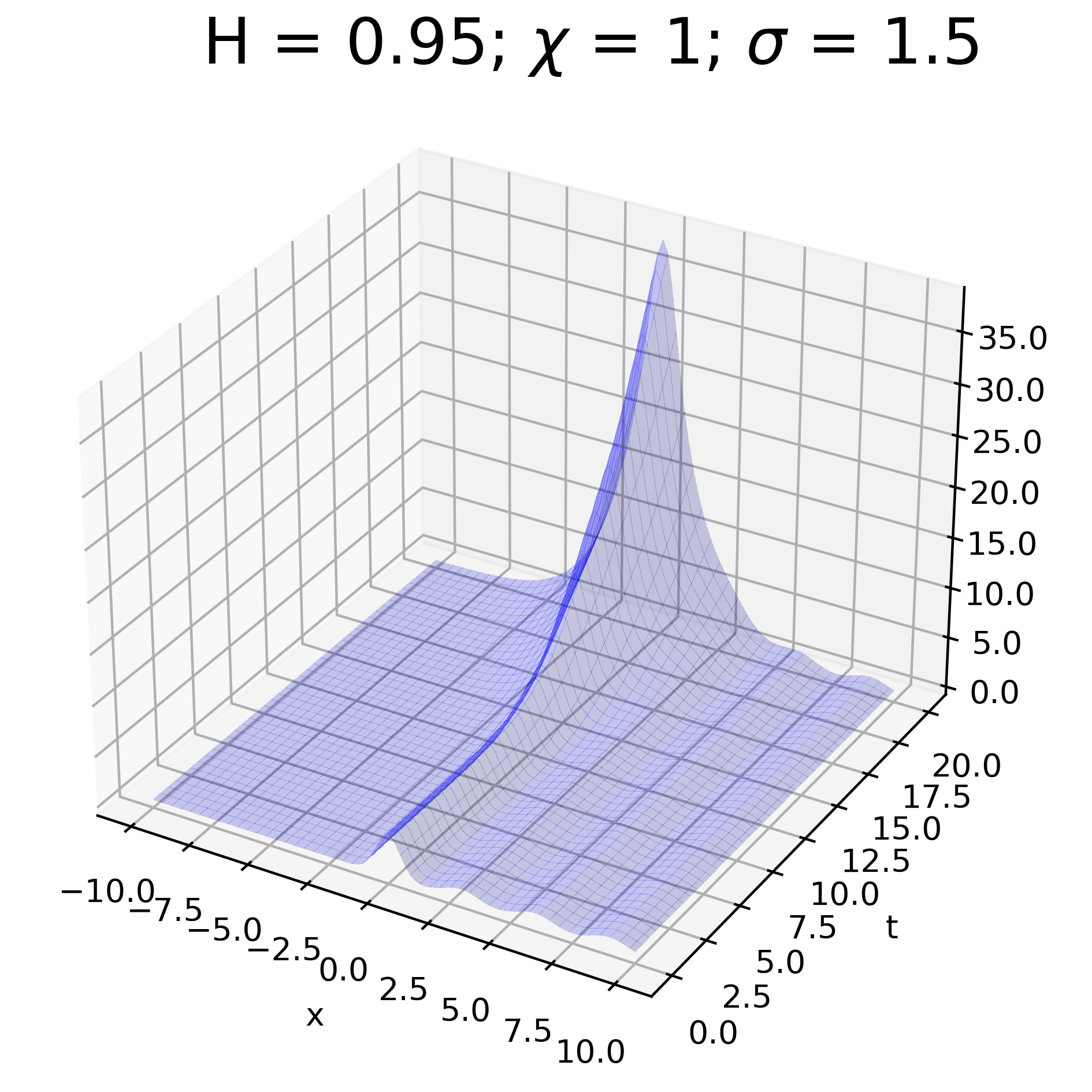}}
    \caption{Time evolution of the combined chemoattractant field $f+h$ for different values of the Hurst index $H$ in the {\it strongly} trapping experiment. }
    \label{fig17}
\end{figure}

\subsection{Influence of secondary self-generated cues}\label{section:3.3}

The results in Sections~\ref{res:Euclid} and \ref{res:man} focused on scenarios where cells (or computational particles in the simulations) move independently under the influence of a primary chemotactic field. In many biological systems, however, chemotactic migration is not purely individual but is reinforced through intercellular communication \cite{chem:2016}. Experimental studies have shown that once cells reach a sufficiently strong primary chemoattractant source, they may secrete secondary signals that reshape the local chemical landscape and recruit additional cells toward the source. This \emph{relay mechanism} effectively transforms the search process from one driven by independent explorers into a collective migration strategy. 

Several well-studied examples highlight this principle. In \emph{Dictyostelium discoideum}, starving cells migrate collectively by relaying waves of cyclic adenosine monophosphate (cAMP): cells both sense external cAMP and release additional cAMP once their receptors are sufficiently activated \cite{Devreotes1989,Kessin2001}. Similarly, neutrophils responding to primary cues such as fMLP, a prototypical bacterial-derived peptide chemoattractant, or tissue damage secrete leukotriene B4 (LTB4), which acts as a secondary chemoattractant that sustains and coordinates the swarming response at inflammatory sites \cite{Afonso2012,Majumdar2021,Strickland2024}. These systems demonstrate how secondary cues can amplify weak or localized signals, ensuring that the population as a whole converges efficiently toward biologically relevant targets.

Motivated by this evidence, we extend our model to incorporate the secretion of a secondary signal by ``activated'' cells. Specifically, we investigate how the interplay between fractional Brownian motion, gradient ascent in the primary field, and secondary signal production alters the collective dynamics. This shifts the framework from independent particles to an interacting system, where trajectories are coupled through the evolving secondary chemoattractant field. 

In our simulations, $M$ computational particles (cells) are initialized in the domain. In addition to the primary chemotactic signal $f$, we introduce a secondary diffusible signal $h$, produced by cells located in regions where the primary chemoattractant exceeds a threshold $r$ ($r=3.5$ in all simulations). Each cell, whether activated or not, migrates along the gradient of the combined field $f+h$, according to:\medskip
\begin{equation}\label{eq331}
\left\{\begin{array}{l}
\displaystyle d_{}\mathbf{X}_t^{k} = \Bigl(\chi\,\nabla f\bigl(\mathbf{X}^k_t\bigr) + \chi\,\nabla h\bigl(\mathbf{X}^k_t,t\bigr)\Bigr)\,dt + \sigma\, d_{}\mathbf{W}_t^{H,k}\medskip\\
\displaystyle\phantom{d_{}} \mathbf{X}^k_0 = \mathbf{x}^k_0
\end{array}\right.
\end{equation}
where $\mathbf{X}^k_t$ denotes the position of the $k$-th cell at time $t$, and $\mathbf{W}_t^{H,k}$ are independent fractional Brownian motions with common Hurst index $H \geq 1/2$ for $k=1,\ldots,M$.  

The secondary signal $h$ evolves according to:\medskip
\begin{equation}\label{eq332}
\left\{\begin{array}{l}
\displaystyle \frac{\partial}{\partial t}h(\mathbf{x},t) = D\,\Delta h(\mathbf{x},t) + \sum_{k=1}^M \mathbbm{1}\Bigl\{ f\bigl(\mathbf{X}_t^{k}\bigr)\geq r \Bigr\}\, G\bigl(\mathbf{x}-\mathbf{X}_t^{k}\bigr) \quad \text{in }\Omega\times\mathbb{R}^+\medskip\\
h(\mathbf{x},t) = 0 \quad \text{on }\partial \Omega\times\mathbb{R}^+
\end{array}\right.
\end{equation}
where $\Omega$ is a smooth domain, $D>0$ is a diffusion constant, and $G(\mathbf{x})$ is a smooth mollification of the Dirac distribution with compact support; see, e.g., Chapter 6 in \cite{Sacks:2017}. In this way, activated cells generate a secondary field that diffuses through the environment and influences remote cells, shifting the system from independent trajectories to coordinated, population-level dynamics. Moreover, throughout this section, we assume that 
\begin{equation}\label{eq333}
h(\mathbf{x},0) = 0\quad\text{in }\Omega
\end{equation}

Figure~\ref{fig17} illustrates the evolution of the combined field $f+h$ in the one-dimensional, strongly trapping experiment. By the initial condition in Eq.~\eqref{eq333}, the combined field coincides with the stationary primary signal $f$ at early times. Only after some trajectories reach the vicinity of the global maximum and initiate secretion of the secondary chemoattractant does $f+h$ begin to deviate from $f$, signaling the onset of interacting dynamics. The panels highlight the influence of the Hurst index $H$. Over the interval $[0,20]$, the combined field remains nearly unchanged for $H=1/2$, whereas larger values of $H$ accelerate the first encounters with the primary chemoattractant source and induce a more rapid rise of the diffusible signal $h$, sharpening the effective landscape near the global maximum.

\begin{figure}[!t]
    \centering
    \includegraphics[width=0.9\textwidth]{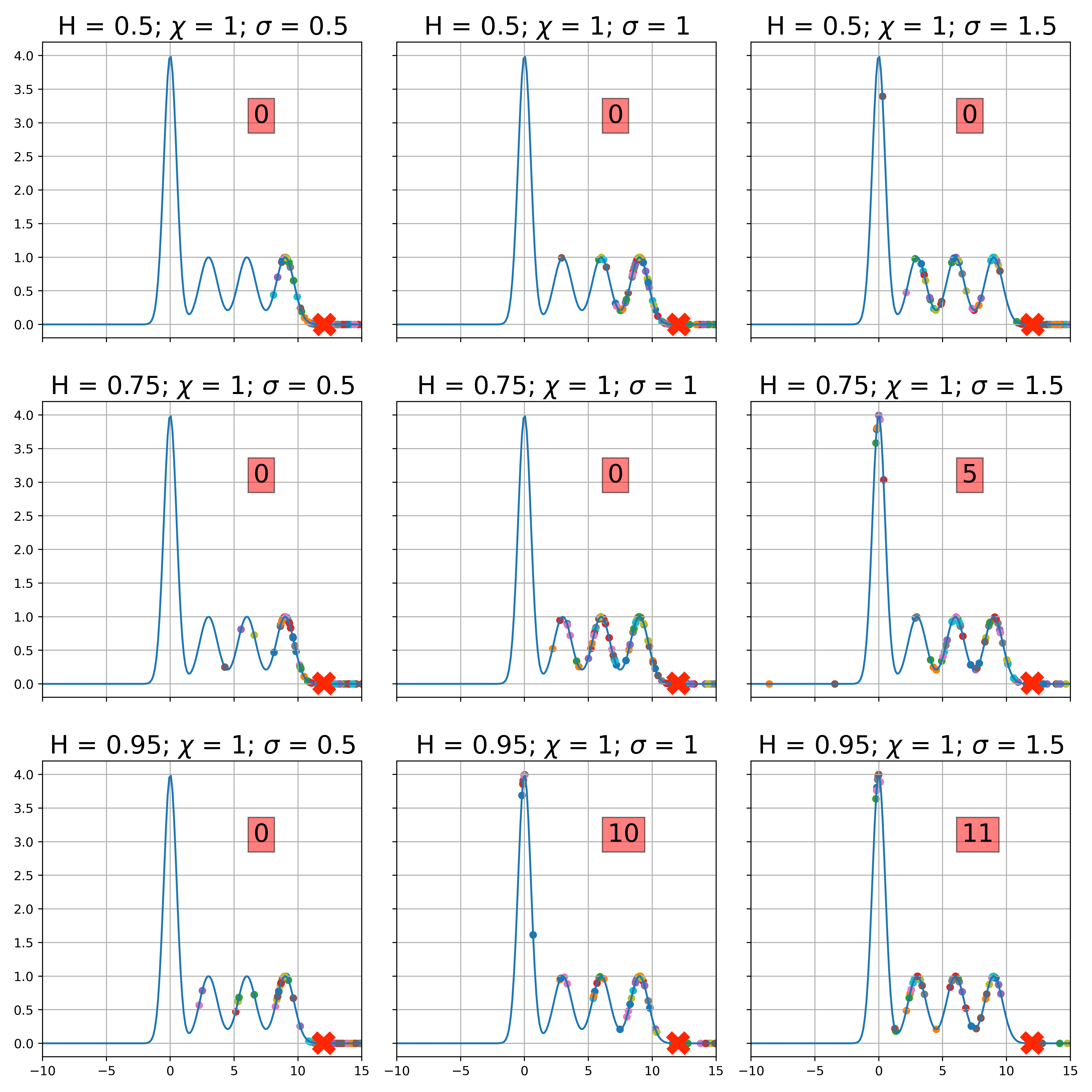}
    \caption{Spatial positions of $100$ interacting particles in the presence of a secondary, diffusible signal $h$ at $t=10$ for different parameter sets in the \emph{strongly} trapping experiment. When compared with the results in Fig.~\ref{fig4}, one observes a swarming response at the site of the global maximum of the primary signal $f$. }
    \label{fig18}
\end{figure}

Figure~\ref{fig18} shows the positions of $100$ interacting particles at $t=10$ in the strongly trapping regime. Compared with the independent case (Fig.~\ref{fig4}), two effects are evident. First, the robustness conferred by larger Hurst indices persists, as temporal correlations continue to facilitate escape from local maxima. Second, even though the secondary field has only just begun to rise by $t=10$ (as seen in Fig.~\ref{fig17}), it already enhances recruitment to the global maximum of $f$, producing an incipient swarming response and reducing occupancy of distant peaks.

\begin{figure}[!t]
    \centering
    \hspace*{-0.7cm}\includegraphics[width=0.98\textwidth]{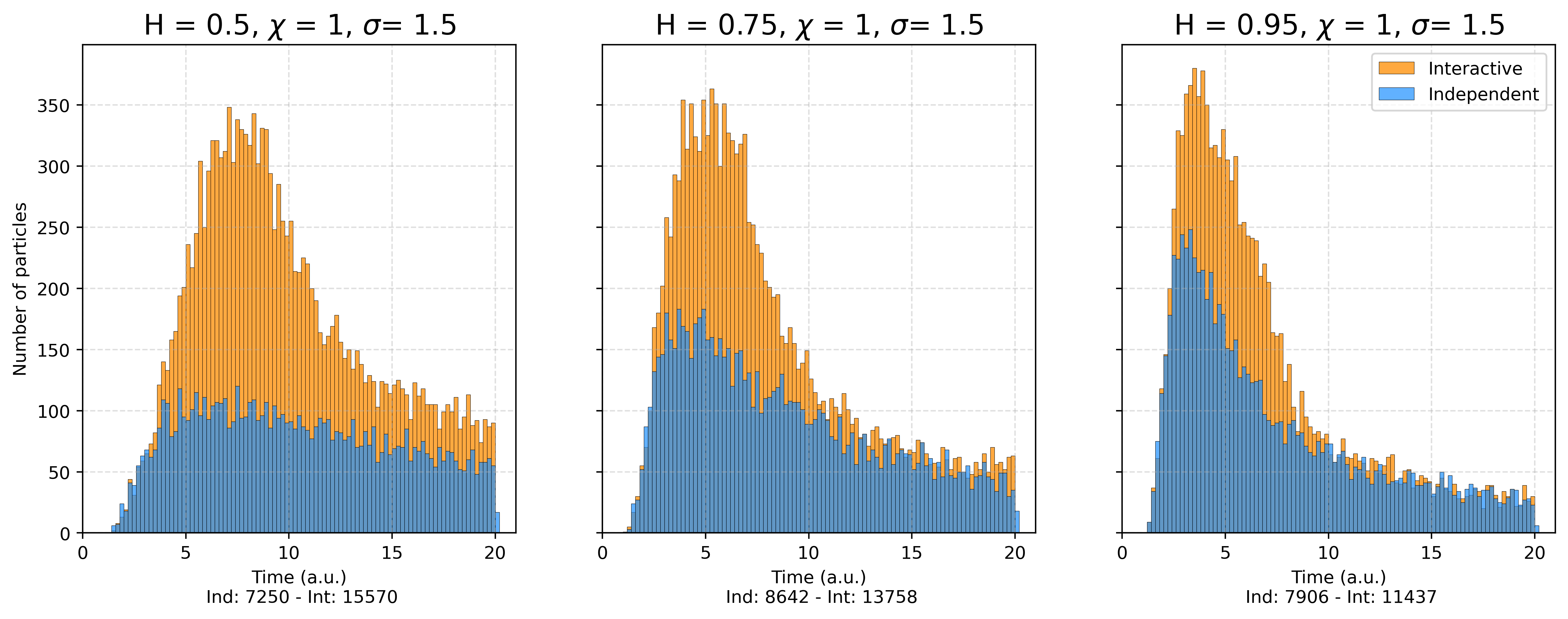}
    \caption{Comparison of the histograms for the conditional hitting time distributions in Fig.~\ref{fig7} (in blue) with the corresponding histograms for the interacting-particles system (in orange) in the  {\it weakly} trapping experiment.}
    \label{fig19}
\end{figure}

\begin{figure}[!t]
    \centering
    \hspace*{-0.7cm}\includegraphics[width=0.98\textwidth]{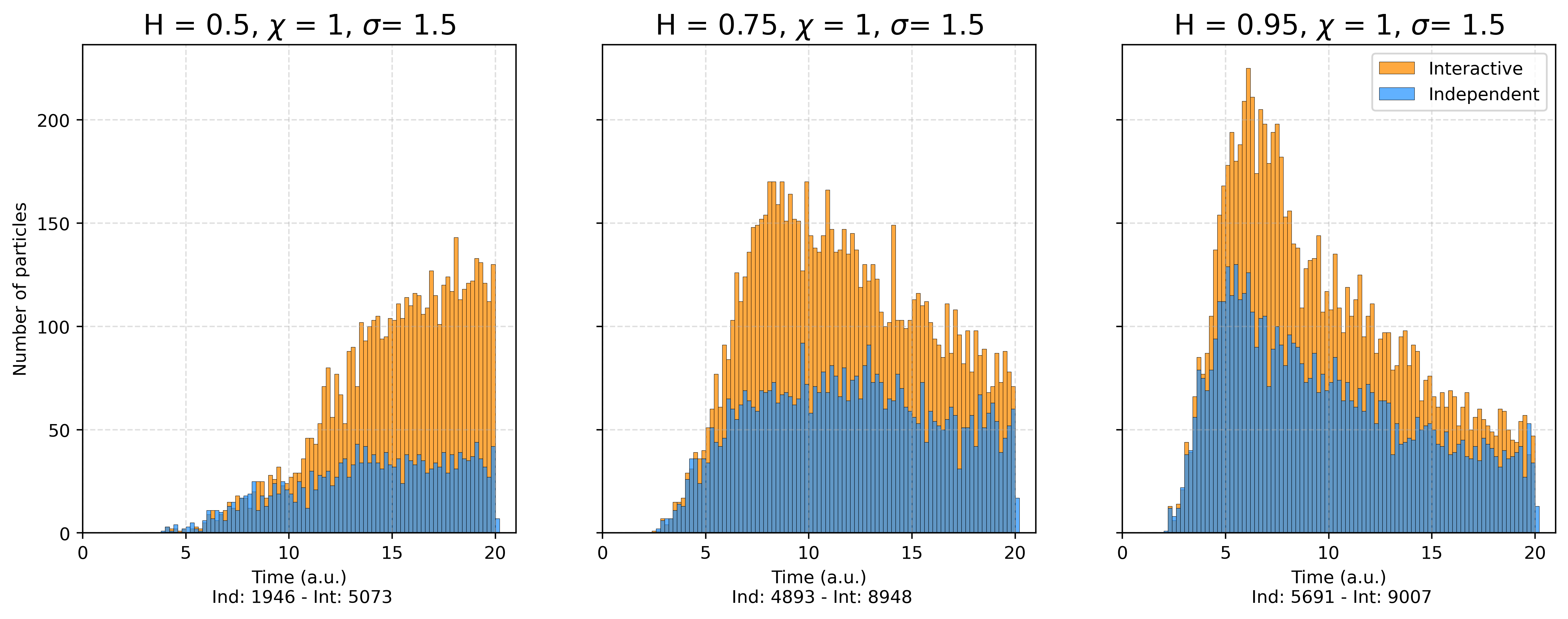}
    \caption{Comparison of the histograms for the conditional hitting time distributions in Fig.~\ref{fig8} (in blue) with the corresponding histograms for the interacting-particles system (in orange) in the  {\it strongly} trapping experiment.}
    \label{fig20}
\end{figure}

Figures~\ref{fig19} and \ref{fig20} compare the empirical hitting-time distributions for the interacting system (orange) with those of the independent baseline (blue) in the weakly and strongly trapping regimes, respectively. In both cases, the presence of the secondary cue shifts probability mass toward shorter hitting times and increases the fraction of trajectories that reach the global maximum within the observation window. The labels \texttt{Ind} and \texttt{Int} beneath the panels denote the total number of successful trajectories in the independent- and interacting-particle simulations, respectively, providing a direct measure of the recruitment effect induced by secondary signaling.

An additional distinction from the independent-particle simulations of Sections~\ref{res:Euclid} and \ref{res:man} is that, in the interacting framework, cells do not terminate their motility upon reaching regions where the primary signal exceeds the threshold $r$. Instead, they continue their chemotactic program, where the secondary signal $h$ plays a central role in maintaining cells within the vicinity of the global maximum of $f$.

\begin{figure}[!t]
    \centering
    \includegraphics[width=1\textwidth]{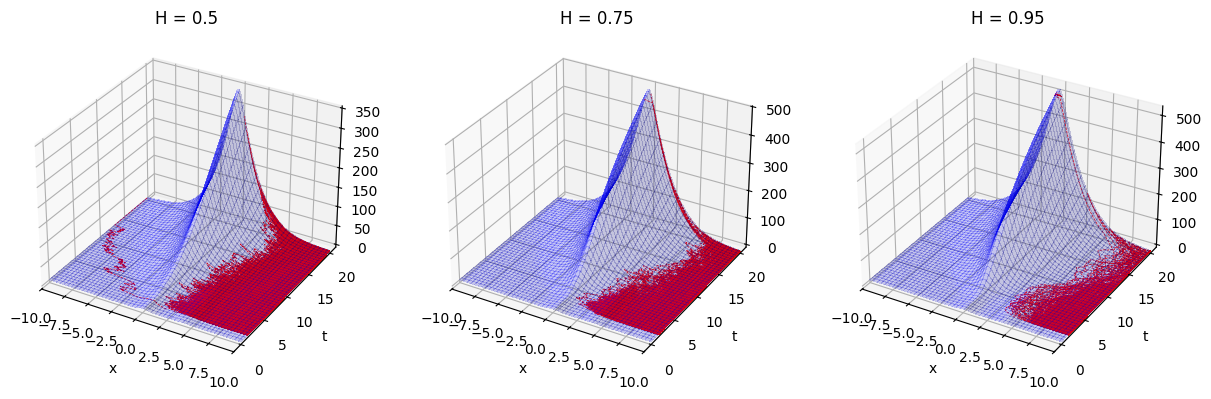}
    \caption{Time evolution of the combined chemoattractant field $f+h$ for different values of the Hurst index $H$ in the {\it weakly} trapping experiment. The trajectories in red correspond to sample paths for which $f(X_\tau)<r$ at  $\tau=20$ (see text for more details).}
    \label{fig21}
\end{figure}

Figure~\ref{fig21} depicts trajectories that fail to reach the target during the observation interval in the weakly trapping regime. As the panels indicate, this population progressively recedes as the front of the secondary signal advances. The left panel highlights a solitary trajectory that escapes from the global maximum despite the stabilizing influence of $h$, indicating that while secondary signaling greatly enhances retention, it does not entirely preclude escape.

\section{Discussion}

\FloatBarrier
\FloatBarrier

The computational experiments presented in this paper provide evidence that temporal correlations in random cell motility, modeled here through fractional Brownian motion, play a functional role in chemotactic navigation. Across Euclidean domains, curved substrates, and scenarios with secondary self-generated cues, superdiffusive motion consistently enhanced the ability of at least one trajectory to reach the global maximum of the primary chemoattractant field. This robustness was observed despite environmental fluctuations, competing local maxima, or geometric constraints, suggesting that memory effects in random motility can improve search efficiency in biologically relevant landscapes.

Our analysis of independent particles in Euclidean domains showed that increasing the Hurst index $H$ accelerates escape from local maxima and raises the likelihood that a particle reaches the global maximum within a fixed observation window. These findings are consistent with experimental observations of superdiffusive behavior in fibroblasts, epithelial cells, neutrophils, and other systems, where temporal correlations in motility have been linked to effective exploration of complex environments. In a closely related study of neutrophil chemotaxis, \cite{Dieterich2022} reported persistent anomalous motion of wildtype neutrophils in CXCL1 gradients, whereas TRPC6-knockout or CXCR2-blocked cells exhibited tempered memory and impaired chemotactic performance. Their first-passage simulations further showed accelerated target discovery by wildtype neutrophils relative to the perturbed conditions, consistent with our finding that persistent motion can enhance finite-time source-finding. While that study focused on experimentally observed neutrophil trajectories in a specific chemotactic setting, the present work isolates the role of fractional-Brownian persistence in controlled chemoattractant landscapes, including local trapping regions, curved geometries, and interacting populations with secondary self-generated cues. At the same time, strong persistence may occasionally delay arrival, either by reinforcing motion away from the source or by sustaining unusually small increments along the correct direction. Nevertheless, at the population level, the dominant effect of long-range correlations was an overall acceleration of the search process, and a minimal one-dimensional calculation illustrating this finite-window conditional first-passage effect is given in Appendix 4.

\begin{figure}[!t]
    \centering
    \includegraphics[width=1.0\textwidth]{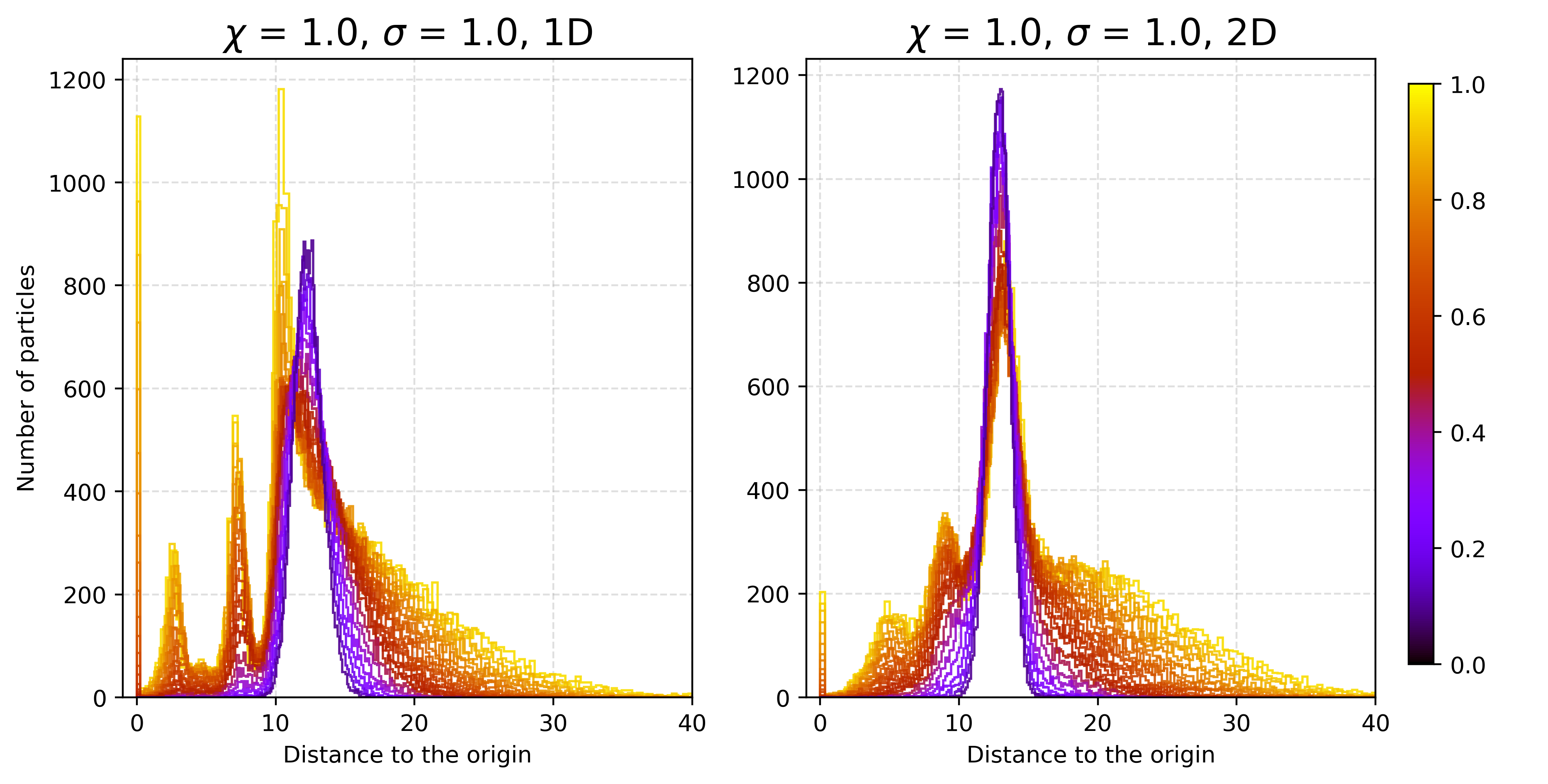}
    \caption{Distribution of particle distances from the origin for the one-dimensional strongly trapping experiment (left) and its two-dimensional counterpart (right), shown for various Hurst indices (color-coded in the colorbar on the left). Both experiments were performed using $20,000$ particles, and the distributions are shown at time $t=20$. The enhanced localization of particles for $H<1/2$ is clearly apparent.}
    \label{fig22}
\end{figure}

For this reason, the present study focused on the superdiffusive regime $H>1/2$, where temporal correlations enhance exploration and improve finite-time source-finding. By contrast, when $H<1/2$, the mean-squared displacement of fractional Brownian motion grows sublinearly in time, and therefore more slowly than in the Brownian case $H=1/2$. Thus, one expects reduced exploratory capacity, at least at the level of spatial spread. This is confirmed in Fig.~\ref{fig22}, which shows the fixed-time distribution of particle distances from the origin in both the one-dimensional strongly trapping experiment and its two-dimensional analogue. In both cases, the distributions for $H<1/2$ are more sharply localized than their Brownian and superdiffusive counterparts, indicating that anti-persistent increments tend to confine the bulk of the population near the trapping region rather than promote global exploration. This conclusion concerns the typical finite-time spatial spread of the ensemble and does not preclude distinct extreme-value effects, such as rare early arrivals in large ensembles of the type investigated in \cite{Lawley:2026}. The localization observed in Fig.~\ref{fig22} is also consistent with the experimental and modeling literature, where subdiffusive single-cell motility is often associated with confinement, adhesion, crowding, or other mechanisms that reduce effective exploration and promote more localized motion \cite{Belotti,Luzhansky,Ravasio:2026}.

The same fractional framework is also relevant beyond cell-level chemotactic exploration. In particular, there is substantial experimental evidence that anomalous intracellular motion in crowded cytoplasmic or nuclear environments can be described in terms of fractional Brownian motion, often in the subdiffusive regime. For example, Burnecki \textit{et al.} \cite{Burnecki} provided statistical evidence supporting a fractional Brownian motion description of subdiffusive telomere motion in living human U2OS cells, while Sabri \textit{et al.} \cite{Sabri} reported microtubule-dependent subdiffusion with antipersistent increments for tracer motion in the cytoplasm of living mammalian cells, modeled in terms of intermittent fractional Brownian motion. More generally, Weiss \cite{WeissM} discussed how crowded viscoelastic environments give rise to the signatures of anticorrelated fractional Brownian motion. These works suggest that fractional stochastic models of the type considered here may also be useful for describing intracellular transport and the motion of biomolecules in the cytoplasm and nucleus.

At the cellular level, the clearest experimental evidence for fractional-Brownian-motion-type behavior in migration remains the literature discussed in Section \ref{intro}, including fibroblasts, epithelial cells, neutrophils, and \textit{Drosophila} hemocytes. At the same time, the biological relevance of the present framework likely extends more broadly. In particular, recent work has modeled serotonergic axons as sample paths of reflected fractional Brownian motion, showing that such dynamics can account for aspects of the spatial organization of fiber densities in the brain \cite{Janusonis:2020,Janusonis:2025}. Related studies of axonal growth in controlled geometries have likewise emphasized the interplay between deterministic guidance and stochastic exploration, with anomalous superdiffusive behavior emerging at longer times \cite{Yurchenko:2019}. More generally, correlated exploratory dynamics of the kind studied here may also prove useful in the modeling of developmental processes, wound healing, cancer invasion, and angiogenesis, where directed migration, environmental structure, and cell--cell or cell--matrix interactions all play central roles; see, for example, \cite{Chaplain:2006,Chaplain:2000,Olsen:1996}. From this perspective, the present work may be viewed as part of a broader effort to understand how temporally correlated stochastic motion influences biologically meaningful search, transport, and invasion processes across multiple scales.

When substrate geometry was introduced, the interplay between curvature-induced biases and temporally correlated noise revealed a similar pattern. On paraboloid substrates, geometric effects promoted circumferential motion that hindered progress toward the source, yet higher values of $H$ mitigated this constraint, allowing trajectories to cross level sets and approach the global maximum. This demonstrates that persistence not only confers robustness against noisy chemotactic fields but also counterbalances directional biases imposed by geometry, broadening the relevance of our findings to settings where topography and curvature influence cell behavior.

Finally, incorporating secondary chemotactic cues transformed the system from one of independent explorers to a coupled population. Once a subset of trajectories reached the source and secreted the secondary signal, recruitment of additional particles accelerated, leading to swarming-like behavior. Comparison of hitting-time distributions showed that this cooperative mechanism amplifies the benefits of temporal correlations, shifting probability mass toward shorter times and increasing the fraction of successful trajectories. These results align with well-established biological relay mechanisms, such as cAMP signaling in \textit{Dictyostelium} and LTB4-mediated swarming in neutrophils, and underscore how cell populations can exploit both intrinsic motility correlations and intercellular communication to achieve robust collective navigation. Although these systems rely on genuinely spatio-temporal chemotactic fields in {\it vivo}, \textit{Dictyostelium} chemotaxis has also been investigated experimentally in controlled {\it in vitro} settings with stationary or near-stationary gradients~\cite{Amselem:2012}.

Beyond biological implications, these findings also have relevance for optimization and sampling. In the Brownian case $H=1/2$, and after identifying the chemoattractant landscape with an objective function to be maximized, or equivalently its negative with a loss function to be minimized, the dynamics considered here reduce to a Brownian-noise gradient system. Such equations are closely related to diffusion approximations of stochastic gradient descent. In particular, \cite{HuLiu:2019} rigorously derived diffusion approximations for nonconvex stochastic gradient descent and used them to analyze the time scales on which Brownian gradient noise drives escape from local minima and saddle points \cite{HuLiu:2019}. This connection places the present results in dialogue with a broader optimization literature in which stochasticity helps trajectories avoid trapping by suboptimal critical points. It is also naturally related to particle swarm optimization, introduced by Kennedy and Eberhart as a stochastic population-based optimization method inspired by collective social behavior, and later developed into a flexible methodology for continuous optimization \cite{Bonyadi:2017,Kennedy:1995}.

A complementary line of work introduces temporal dependence in optimization dynamics through Markovian or asynchronous mechanisms rather than through fractional noise. For example, in the stochastic gradient process of \cite{Latz:2021}, the optimizer follows the gradient flow associated with a sampled component of the objective, and the active component is refreshed only when an underlying continuous-time Markov process jumps. Delayed and asynchronous variants of stochastic gradient descent likewise study algorithms driven by stale or temporally dependent gradient information \cite{Agarwal:2011,Deng:2025}. Although these models differ mathematically from the present one, since they remain Markovian or delay-driven whereas fractional Brownian motion with $H>1/2$ has positively correlated increments and long memory, they reflect a related design principle: persistence in the stochastic driving signal can alter finite-time optimization behavior and may promote escape from trapping regions. In our setting, this persistence is generated by the positive correlations of fractional Brownian motion. Our emphasis in this paper has been on understanding how temporal correlations of biological origin affect search efficiency, trapping, and localization in the superdiffusive regime $H>1/2$. Together, these observations suggest that structured stochasticity may be leveraged not only in cell biology but also in computational settings where robustness to noise and efficiency of exploration are critical.

\subsection*{Appendix 1: Fractional Brownian motion in local coordinates}

In this appendix, we derive Eq.~\eqref{SDE:local} in local coordinates under the assumptions $\chi=0$, $\sigma=1$, and $H>1/2$. The general expression in Eq.~\eqref{SDE:local} follows from this special case together with the expression for the intrinsic gradient in Section~\ref{manifold}. Throughout, all stochastic differentials are understood in the \emph{Young sense} \cite{Nourdin:2012}.

Let $\mathcal{M}\subset\mathbb{R}^N$ be a smooth embedded $n$-dimensional manifold admitting a global chart
$$
\varphi:\;U\subset\mathbb{R}^n \;\longrightarrow\; \mathcal{M}\subset\mathbb{R}^N,\qquad 
\mathbf{x}\longmapsto \mathbf{y}=\varphi(\mathbf{x}).
$$
We denote by 
$$
\partial_j\varphi(\mathbf{x}) \;=\; \frac{\partial\varphi}{\partial x^j}(\mathbf{x}), 
\qquad j=1,\dots,n,
$$
the coordinate tangent vectors, and equip $U$ with the induced Riemannian metric
$$
g_{jk}(\mathbf{x})
\;=\;
\bigl\langle \partial_j\varphi(\mathbf{x}),\,\partial_k\varphi(\mathbf{x}) \bigr\rangle_{\mathbb{R}^N}.
$$
Let $g^{ij}(\mathbf{x})$ denote the components of the inverse matrix $g^{-1}(\mathbf{x})$, and let $g^{-1/2}(\mathbf{x})$ denote the symmetric square root of $g^{-1}(\mathbf{x})$.

Consider an $n$-dimensional \emph{Euclidean} fractional Brownian motion  $\mathbf{W}_t^H$ with independent components and Hurst index $H>1/2$. We define an orthonormal frame along $\mathcal{M}$ by
\begin{equation}\label{frame}
E_k\bigl(\varphi(\mathbf{x})\bigr)
\;=\;
\sum_{j=1}^n 
\partial_j\varphi(\mathbf{x})\,
\bigl(g^{-1/2}(\mathbf{x})\bigr)^{j}_{k},
\qquad k=1,\dots,n.
\end{equation}
Each column $E_k(\varphi(\mathbf{x}))$ is a tangent vector in $T_{\varphi(\mathbf{x})}\mathcal{M}$, and 
$\{E_k(\varphi(\mathbf{x}))\}_{k=1}^n$ forms an orthonormal basis of the tangent space. 
We denote by 
$$
E\bigl(\varphi(\mathbf{x})\bigr)
\;=\;
\bigl[E_1\bigl(\varphi(\mathbf{x})\bigr)\,\cdots\,E_n\bigl(\varphi(\mathbf{x})\bigr)\bigr]
$$
the $N\times n$ matrix whose $k$-th column is $E_k(\varphi(\mathbf{x}))$. 
By construction, $E(\varphi(\mathbf{x}))$ satisfies
$$
E\bigl(\varphi(\mathbf{x})\bigr)^{\!\top}E\bigl(\varphi(\mathbf{x})\bigr)=I_n,
\qquad
E\bigl(\varphi(\mathbf{x})\bigr)E\bigl(\varphi(\mathbf{x})\bigr)^{\!\top}
= P\bigl(\varphi(\mathbf{x})\bigr),
$$
where $P(\varphi(\mathbf{x}))$ is the orthogonal projection onto the tangent space $T_{\varphi(\mathbf{x})}\mathcal{M}$.

We then define the fractional Brownian motion $\mathbf{B}^H_t$ on $\mathcal{M}$ as the solution to the Young stochastic differential equation
\begin{equation}\label{B_def}
\left\{\begin{array}{l}
\displaystyle d_{}\mathbf{B}^H_t \,= \,  \sum_{k=1}^n E_k\bigl(\mathbf{B}^H_t\bigr)\, dW_t^{H,k}\bigskip\\
\phantom{d_{}}\mathbf{B}^H_0\,=\,\boldsymbol{\beta}^H_0\in \mathcal{M}
\end{array}
\right.
\end{equation}
Note that by construction, $d_{}\mathbf{B}^H_t\in T_{\mathbf{B}^H_t}\mathcal{M}$, and therefore $\mathbf{B}^H_t\in\mathcal{M}$ for all $t\ge0$.

Now, define the local coordinate process
\begin{equation}\label{X_def}
\mathbf{X}_t \;=\; \varphi^{-1}\!\bigl(\mathbf{B}^H_t\bigr)\in U.
\end{equation}
Applying Young’s chain rule (see, e.g., \cite{Mishura:2017}) to \eqref{X_def} yields
$$
d_{}\mathbf{X}_t
\;=\;
D\bigl(\varphi^{-1}\bigr)\!\bigl(\mathbf{B}^H_t\bigr)\; d_{}\mathbf{B}^H_t,
$$
where $D(\varphi^{-1})(\mathbf{y})$ is the $n\times N$ Jacobian matrix of $\varphi^{-1}$ at $\mathbf{y}\in\mathcal{M}$. Substituting  \eqref{B_def} gives
$$
d_{}\mathbf{X}_t
\;=\;
D\bigl(\varphi^{-1}\bigr)\!\bigl(\mathbf{B}^H_t\bigr)
\,E\bigl(\mathbf{B}^H_t\bigr)\,
d_{}\mathbf{W}^H_t.
$$

To compute $D(\varphi^{-1})\,E$ explicitly, we note that at any $\mathbf{x}\in U$,
the Jacobian of $\varphi$ satisfies
$$
D\varphi(\mathbf{x})
=
\bigl[\partial_1\varphi(\mathbf{x})\,\cdots\,\partial_n\varphi(\mathbf{x})\bigr],
$$
and the induced metric satisfies $g(\mathbf{x}) = D\varphi(\mathbf{x})^{\!\top}D\varphi(\mathbf{x})$.  
Differentiating the identity $\varphi^{-1}(\varphi(\mathbf{x}))=\mathbf{x}$ gives
$$
D\bigl(\varphi^{-1}\bigr)\!\bigl(\varphi(\mathbf{x})\bigr)\,D\varphi(\mathbf{x}) = I_n.
$$
Multiplying both sides on the right by $D\varphi(\mathbf{x})^{\!\top}$ and rearranging yields
$$
D\bigl(\varphi^{-1}\bigr)\!\bigl(\varphi(\mathbf{x})\bigr)
= g^{-1}(\mathbf{x})\,D\varphi(\mathbf{x})^{\!\top}.
$$
Substituting $E(\varphi(\mathbf{x})) = D\varphi(\mathbf{x})\,g^{-1/2}(\mathbf{x})$, we obtain
$$
D\bigl(\varphi^{-1}\bigr)\!\bigl(\varphi(\mathbf{x})\bigr)\,E\bigl(\varphi(\mathbf{x})\bigr)
= g^{-1}(\mathbf{x})\,D\varphi(\mathbf{x})^{\!\top}D\varphi(\mathbf{x})\,g^{-1/2}(\mathbf{x})
= g^{-1/2}(\mathbf{x}),
$$
where the last equality follows from the definition of the metric $g = D\varphi^{\!\top}D\varphi$.

Finally, evaluating at $\mathbf{x}=\mathbf{X}_t$ gives the desired local coordinate formulation:
$$
d_{}\mathbf{X}_t
\;=\;
g^{-1/2}\bigl(\mathbf{X}_t\bigr)\, d_{}\mathbf{W}^H_t,
$$
or, componentwise,
\begin{equation}\label{SDE_local_final}
d_{}X_t^{\,i}
\;=\;
\sum_{k=1}^n 
\bigl(g^{-1/2}\bigr)^{i}_{k}\bigl(\mathbf{X}_t\bigr)\, d_{}W_t^{H,k},
\qquad i=1,\dots,n.
\end{equation}
Equation~\eqref{SDE_local_final} coincides with Eq.~\eqref{SDE:local} for the special case $\chi=0$ and $\sigma=1$, and is valid in the Young sense for all $H>1/2$.

\subsection*{Appendix 2: The inverse square root of the metric tensor}

The expressions for the geometric tensors in Section~\ref{res:man} can be readily verified. Regarding Eq.~\eqref{sqinvm} as an example,  we have:

\begin{align*} \sum_{k=1}^n\left(g^{-1/2}\right)^i_k\left(g^{-1/2}\right)^k_j&=\sum_{k=1}^n\left(\delta^i_k-\frac{\partial^i \psi\,\partial_k\psi}{\operatorname{det}(g)+\sqrt{\operatorname{det}(g)}}\right)\left(\delta^k_j-\frac{\partial^k \psi\,\partial_j \psi}{\operatorname{det}(g)+\sqrt{\operatorname{det}(g)}}\right)
\\ &= \sum_{k=1}^n\delta^i_k\delta^k_j - \sum_{k=1}^n\delta^i_k\frac{\partial^k \psi\,\partial_j \psi}{\operatorname{det}(g)+\sqrt{\operatorname{det}(g)}} 
\\& \phantom{=\;} -\sum_{k=1}^n\delta^k_j\frac{\partial^i \psi\,\partial_k\psi}{\operatorname{det}(g)+\sqrt{\operatorname{det}(g)}}
+ \sum_{k=1}^n\frac{\partial^i \psi\,\partial_k\psi\partial^k \psi\,\partial_j \psi}{\Bigl(\operatorname{det}(g)+\sqrt{\operatorname{det}(g)}_{}\Bigr)^2}
\\ &= \delta^{ij} - \frac{2\partial^i \psi\,\partial^j \psi}{\operatorname{det}(g)+\sqrt{\operatorname{det}(g)}} +\frac{(\operatorname{det}(g)-1)\partial^i \psi\,\partial^j \psi}{\displaystyle\Bigl(\operatorname{det}(g)+\sqrt{\operatorname{det}(g)}_{}\Bigr)^2}
\\ &= \delta^{ij} + \left(\frac{\operatorname{det}(g)-1}{(\operatorname{det}(g)+\sqrt{\operatorname{det}(g)})^2}-\frac{2}{\operatorname{det}(g)+\sqrt{\operatorname{det}(g)}} \right)\partial^i \psi\,\partial^j \psi
\\ &=\delta^{ij} + \left(\frac{(\sqrt{\operatorname{det}(g)} -1)(\sqrt{\operatorname{det}(g)} + 1)}{\operatorname{det}(g)(\sqrt{\operatorname{det}(g)}+1)^2}-\frac{2}{\sqrt{\operatorname{det}(g)}(\sqrt{\operatorname{det}(g)}+1)} \right)\partial^i \psi\,\partial^j \psi
\\ &= \delta^{ij} + \left(\frac{\sqrt{\operatorname{det}(g)} -1}{\operatorname{det}(g)(\sqrt{\operatorname{det}(g)}+1)}-\frac{2}{\sqrt{\operatorname{det}(g)}(\sqrt{\operatorname{det}(g)}+1)} \right)\partial^i \psi\,\partial^j \psi
\\ &= \delta^{ij} + \left(\frac{\sqrt{\operatorname{det}(g)} -1 - 2\sqrt{\operatorname{det}(g)}}{\operatorname{det}(g)(\sqrt{\operatorname{det}(g)}+1)}\right)\partial^i \psi\,\partial^j \psi
\\ &=\delta^{ij} - \left(\frac{\sqrt{\operatorname{det}(g)} +1}{\operatorname{det}(g)(\sqrt{\operatorname{det}(g)}+1)}\right)\partial^i \psi\,\partial^j \psi
\\ &= \delta^{ij} - \frac{\partial^i \psi\,\partial^j \psi}{\operatorname{det}(g)} = g^{ij}
\end{align*}

\subsection*{Appendix 3: The stationary chemoattractant profiles}

In this appendix, we provide the analytic expressions for the stationary chemoattractant profiles that appear in Section \ref{res:Euclid}. The stationary profile associated with the experiment in Fig.~\ref{fig2} is given by the function
\begin{align}
    f_{\text{weak}}(x) = 4\,e^{-2x^2} + e^{-(x-4)^2}\nonumber
\end{align}
The stationary profile associated with the experiment in Fig.~\ref{fig4} is given by
\begin{align}
    f_{\text{strong}}(x) = 4\,e^{-2x^2} + e^{-(x-3)^2} + e^{-(x-6)^2} + e^{-(x-9)^2}\nonumber
\end{align}
The two-dimensional chemoattractant concentration profile in Fig.~\ref{fig3} is given by
\begin{align}
f_{\text{weak}}(\boldsymbol{x})
=
4 \,e^{
    - \|\boldsymbol{x}\|^2
}
+
e^{
    - \frac{1}{4}\left( \|\boldsymbol{x}\| - 5 \right)^{2}
}\nonumber
\end{align}
The two-dimensional chemoattractant concentration profile in Fig.~\ref{fig5} is given by
\begin{align}
    f_{\text{strong}}(\boldsymbol{x}) = 4 \,e^{
    - \|\boldsymbol{x}\|^2
}
+ e^{
    - \frac{1}{4}\left( \|\boldsymbol{x}\| - 5 \right)^{2}
}
+ e^{
    - \frac{1}{4}\left( \|\boldsymbol{x}\| - 9 \right)^{2}
}
+ e^{
    - \frac{1}{4}\left( \|\boldsymbol{x}\| - 13 \right)^{2}
}\nonumber
\end{align}
Finally, the two-dimensional chemoattractant concentration profiles used on the curved surface in Section \ref{res:man} are the same as above, but defined in the local coordinate system.

\subsection*{Appendix 4: Finite-window conditioning and first-passage scaling}

The hitting-time histograms reported in this work are empirical distributions over trajectories that reach the prescribed target region before the end of the observation window.  They should therefore be interpreted as \emph{conditional} first-passage-time distributions.  The purpose of this appendix is to record the corresponding one-dimensional calculation in a minimal setting, and to isolate the scaling regime in which fractional Brownian motion with $H>1/2$ is expected to hit a target faster than Brownian motion after conditioning on successful detection before a finite time.  We consider a particle initialized at distance $x_0>0$ from the target and write
$$
    \tau=\inf\{t\geq 0:X_t=0\}.
$$
The noise amplitude is set equal to one.  If the stochastic term is instead $\sigma dB_t$ or $\sigma dB_t^H$, then $x_0$ below should be read as the scaled distance $x_0/\sigma$.

For standard Brownian motion, $X_t=x_0+B_t$, the first-passage law follows from the reflection principle; see, for example, \cite{Salminen,Karatzas,Redner:2001}.  The density and distribution function are
\begin{align}
    p_B(t;x_0)
    &=
    \frac{x_0}{\sqrt{2\pi t^3}}
    \exp\!\left(-\frac{x_0^2}{2t}\right),
    \qquad t>0,
    \nonumber\\
    F_B(t;x_0)
    &=
    \mathbb{P}(\tau_B\leq t)
    =
    \operatorname{erfc}\!\left(\frac{x_0}{\sqrt{2t}}\right)
    =
    2\left[1-\Phi\!\left(\frac{x_0}{\sqrt{t}}\right)\right].
    \nonumber
\end{align}
Although $\tau_B<\infty$ almost surely, its unconditioned mean is infinite, since
$$
    \mathbb{P}(\tau_B>t)
    =
    2\Phi\!\left(\frac{x_0}{\sqrt{t}}\right)-1
    \sim
    \sqrt{\frac{2}{\pi}}\,x_0\,t^{-1/2},
    \qquad t\to\infty.
$$
By contrast, conditioning on the finite observation event $\{\tau_B<T_{\rm obs}\}$ gives the truncated and renormalized law
$$
    p_B(t;x_0\mid \tau_B<T_{\rm obs})
    =
    \frac{p_B(t;x_0)}{F_B(T_{\rm obs};x_0)},
    \qquad 0<t<T_{\rm obs},
$$
and hence the finite conditional mean
$$
    m_B(x_0,T_{\rm obs})
    :=
    \mathbb{E}[\tau_B\mid \tau_B<T_{\rm obs}]
    =
    x_0\sqrt{\frac{2T_{\rm obs}}{\pi}}
    \frac{\exp\!\left(-\frac{x_0^2}{2T_{\rm obs}}\right)}
    {\operatorname{erfc}\!\left(\frac{x_0}{\sqrt{2T_{\rm obs}}}\right)}
    -x_0^2 .
$$
This expression gives the two useful limits
\begin{align}
    m_B(x_0,T_{\rm obs})
    &=T_{\rm obs}-\frac{2T_{\rm obs}^2}{x_0^2}
    +o\!\left(\frac{T_{\rm obs}^2}{x_0^2}\right),
    &&T_{\rm obs}\ll x_0^2,
    \nonumber\\
    m_B(x_0,T_{\rm obs})
    &\sim x_0\sqrt{\frac{2T_{\rm obs}}{\pi}},
    &&T_{\rm obs}\gg x_0^2.
    \label{eq:app4_BM_large_window}
\end{align}
Thus, when $T_{\rm obs}$ is short relative to the Brownian scale $x_0^2$, the trajectories that succeed before $T_{\rm obs}$ tend to hit close to the endpoint of the observation window.  For large windows the conditional mean grows like $T_{\rm obs}^{1/2}$ and therefore still diverges as $T_{\rm obs}\to\infty$.

We now consider fractional Brownian motion, $X_t=x_0+B_t^H$, with $H>1/2$ and normalization $\mathbb{E}[(B_t^H)^2]=t^{2H}$.  Fractional Brownian motion is self-similar, $B_{ct}^H\overset{d}{=}c^HB_t^H$, has stationary increments, and has positively correlated increments for $H>1/2$; see \cite{Mishura:2017,VanNess,Nourdin:2012}.  It is not Markovian and the Brownian reflection principle is unavailable.  Nevertheless, with
$$
    M_t^H=\sup_{0\leq s\leq t}B_s^H,
    \qquad
    G_H(y)=\mathbb{P}(M_1^H<y),
$$
symmetry and self-similarity give
\begin{align}
    \mathbb{P}(\tau_H(x_0)\leq t)
    &=\mathbb{P}(M_t^H\geq x_0)
    =1-G_H(x_0t^{-H}),
    \nonumber\\
    \mathbb{P}(\tau_H(x_0)>t)
    &=G_H(x_0t^{-H}),
    \nonumber\\
    \tau_H(x_0)&\overset{d}{=}x_0^{1/H}\tau_H(1).
    \label{eq:app4_fBM_scaling_tau}
\end{align}
The natural fractional Brownian motion  hitting scale is therefore $x_0^{1/H}$, whereas the Brownian scale is $x_0^2$.  If $G_H$ has density $g_H$, then
$$
    p_H(t;x_0\mid \tau_H<T_{\rm obs})
    =
    \frac{Hx_0t^{-H-1}g_H(x_0t^{-H})}
    {1-G_H(x_0T_{\rm obs}^{-H})},
    \qquad 0<t<T_{\rm obs}.
$$
Equivalently, without differentiating the maximum distribution, the conditional mean can be written as
$$
    m_H(x_0,T_{\rm obs})
    =
    \frac{
    \displaystyle\int_0^{T_{\rm obs}}S_H(u;x_0)\,du
    -T_{\rm obs}S_H(T_{\rm obs};x_0)}
    {1-S_H(T_{\rm obs};x_0)},
    \qquad
    S_H(t;x_0)=\mathbb{P}(\tau_H(x_0)>t).
$$
The scaling relation \eqref{eq:app4_fBM_scaling_tau} gives the exact conditional scaling form
$$
    m_H(x_0,T_{\rm obs})
    =
    x_0^{1/H}\,
    \psi_H\!\left(\frac{T_{\rm obs}}{x_0^{1/H}}\right),
    \qquad
    \psi_H(a)=\mathbb{E}[\tau_H(1)\mid \tau_H(1)<a].
$$
For short dimensionless windows, $a\ll 1$, the conditioning is an extreme early-hit event and the conditional law concentrates near the endpoint, giving
$$
    \psi_H(a)\sim a,
    \qquad a\downarrow 0,
    \qquad\text{and hence}\qquad
    m_H(x_0,T_{\rm obs})\sim T_{\rm obs}
    \quad\text{if }T_{\rm obs}\ll x_0^{1/H}.
$$
For large windows, Molchan's persistence theorem gives
\begin{equation}\label{eq:app4_Molchan}
    \mathbb{P}(\tau_H(1)>t)
    =
    t^{-(1-H)+o(1)},
    \qquad t\to\infty,
\end{equation}
see \cite{Simon:2015,Wiese:2015,Ding:1995,Molchan:1999}.  This implies the exponent-level conditional-mean scaling
$$
    \psi_H(a)=a^{H+o(1)},
    \qquad a\to\infty.
$$
The statement \eqref{eq:app4_Molchan} is weaker than a full constant-level asymptotic.  If one additionally assumes regular variation,
$$
    \mathbb{P}(\tau_H(1)>t)=t^{-(1-H)}L_H(t),
$$
with $L_H$ slowly varying (see, e.g., \cite{Taqqu:2017}), then Karamata's theorem gives
$$
    \psi_H(a)
    \sim
    \frac{1-H}{H}\,a\,\mathbb{P}(\tau_H(1)>a)
    =
    \frac{1-H}{H}\,a^H L_H(a).
$$
If, in the special case, $L_H(a)\to C_H$, then
\begin{equation}\label{eq:app4_fBM_large_window_dimensional_constant}
    m_H(x_0,T_{\rm obs})
    \sim
    K_Hx_0^{(1-H)/H}T_{\rm obs}^{H},
    \qquad
    K_H=\frac{1-H}{H}C_H.
\end{equation}
Thus the constant-level formula requires the additional regular-variation assumption, while the power-law exponent $H$ follows from the persistence exponent.  In all cases, the conditional  mean hitting time for fractional Brownian motion is finite for each finite $T_{\rm obs}$ but diverges as the observation window is sent to infinity.

The comparison with Brownian motion is now controlled by the two thresholds
$$
    a_H=\frac{T_{\rm obs}}{x_0^{1/H}},
    \qquad
    a_B=\frac{T_{\rm obs}}{x_0^2}.
$$
Assume $H>1/2$ and $x_0>1$, so that $x_0^{1/H}<x_0^2$.  If
$$
    x_0^{1/H}\ll T_{\rm obs}\ll x_0^2,
$$
then the window is long compared with the fractional Brownian motion scale but short compared with the Brownian scale.  Consequently,
$$
    m_B(x_0,T_{\rm obs})\sim T_{\rm obs},
    \qquad
    m_H(x_0,T_{\rm obs})
    \approx
    T_{\rm obs}
    \left(\frac{x_0^{1/H}}{T_{\rm obs}}\right)^{1-H}
    \ll T_{\rm obs},
$$
where $\approx$ denotes equality at the level of powers, up to constants and slowly varying factors.  Hence
$$
    m_H(x_0,T_{\rm obs})\ll m_B(x_0,T_{\rm obs}),
    \qquad
    x_0^{1/H}\ll T_{\rm obs}\ll x_0^2.
$$
This is a finite-window regime in which fractional Brownian motion is conditionally faster than Brownian motion.

If $T_{\rm obs}\ll x_0^{1/H}<x_0^2$, both processes are conditioned on extremely early hits and both conditional means are asymptotic to $T_{\rm obs}$, so neither process has a meaningful conditional speed advantage.  For very large windows, comparison of \eqref{eq:app4_BM_large_window} with the regular-variation estimate \eqref{eq:app4_fBM_large_window_dimensional_constant} gives, up to constants and slowly varying factors,
$$
    \frac{m_H(x_0,T_{\rm obs})}{m_B(x_0,T_{\rm obs})}
    \asymp
    \left(\frac{T_{\rm obs}}{x_0^{2/H}}\right)^{H-1/2}
$$
Thus fractional Brownian motion  may remain conditionally faster beyond the Brownian scale $x_0^2$, but only up to a larger crossover scale of order $x_0^{2/H}$; for asymptotically very large windows, the heavier  persistence tail eventually makes the fractional Brownian motion  conditional mean larger than the Brownian one.  The robust finite-window conclusion is that fractional Brownian motion is conditionally faster when the observation window is large enough to resolve the superdiffusive scale $x_0^{1/H}$, but still short relative to the Brownian scale $x_0^2$.

\subsection*{Code and data availability}
Code used to generate the simulation data in this study can be found in a public repository at \url{https://github.com/lsbcma/fBmChemotaxis}.

\subsection*{Authors' contributions}
G.C.O. and L.B. developed code, ran simulations, visualized and analyzed data and
drafted the manuscript; J.D., R.E., A.R., and A.M. edited the manuscript and conceived, designed and coordinated the study. 

\subsection*{Competing interests}
The authors declare no competing interests.

\subsection*{Funding}
A.R. was partially supported by ANID FONDECYT Regular 1210872 and 1250073, ANID
FONDEQUIP EMQ210101, and ANID Núcleo Milenio SELFO NCN2024-068. G.C.O., L.B., and  A.M. were partially supported by ANID FONDECYT Regular 1221220.

\bibliographystyle{plain}
\bibliography{main.bib}

\end{document}